%% file: Chiral-Memory.tex
\documentclass[11pt,a4paper]{article}
\usepackage{jheppub}
\usepackage{array}
\usepackage{mathtools}
\usepackage{graphicx}
\usepackage{color}
\usepackage{amsmath}

\newcommand{\be}{\begin{equation}}
\newcommand{\bse}{\begin{subequations}}
\newcommand{\ese}{\end{subequations}}
\newcommand{\bea}{\begin{eqnarray}}
\newcommand{\eea}{\end{eqnarray}}
\newcommand{\ba}{\begin{array}}
\newcommand{\ea}{\end{array}}
\newcommand{\ee}{\end{equation}}

\newcommand{\E}{\boldsymbol{e}}
\newcommand{\hz}{\hat{\boldsymbol{z}}}
\newcommand{\hx}{\hat{\boldsymbol{x}}}
\newcommand{\hy}{\hat{\boldsymbol{y}}}
\newcommand{\bq}{\boldsymbol{q}}
\newcommand{\bdd}{\boldsymbol{d}}
\newcommand{\bmu}{\boldsymbol{\mu}}
\newcommand{\bJ}{\textbf{\textit{J}}}

\newcommand{\bR}{\boldsymbol{R}}

\newcommand{\bDel}{{\boldsymbol{\nabla}}}
\newcommand{\bcJ}{{\bf{J}}}
\newcommand{\bL}{\textbf{\textit{L}}}
\newcommand{\bS}{\textbf{\textit{S}}}
\newcommand{\br}{{\bf{r}}}

\newcommand{\bv}{{\bf{v}}}
\newcommand{\bj}{{\bf{j}}}

\usepackage{xcolor,colortbl}
\definecolor{darkred}{rgb}{0.7,0,0}
\definecolor{purple}{rgb}{0.58,0,0.84}
\definecolor{green}{rgb}{0.07,0.7,0.07}

\def\p{\partial}
\def\bz{{\bar z}}

\newcommand{\mO}{\mathcal{O}}
\newcommand{\mK}{\mathcal{K}}
\newcommand{\bmK}{\boldsymbol{\mathcal{K}}}
\newcommand{\scri}{\mathcal{I}}
\newcommand{\scrip}{\mathcal{I}^+}
\newcommand{\scrim}{\mathcal{I}^-}

\newcommand{\an}{\quad \textmd{and} \quad}

\newcommand{\bn}{{\boldsymbol{n}}}

\newcommand{\bA}{{\bf{A}}}
\newcommand{\bB}{{\bf{B}}}
\newcommand{\bE}{{\bf{E}}}
\newcommand{\bd}{\boldsymbol{\nabla}}
\newcommand{\bh}{{\bf{h}}}

\newcommand{\bX}{{\bf{X}}}
\newcommand{\mA}{\mathcal{A}}
\newcommand{\bk}{{\bf{k}}}
\newcommand{\bx}{{\bf{x}}}

\preprint{CERN-TH-2023-052}

\title{Photon Chiral Memory Effect Stored on Celestial Sphere}
\author{Azadeh Maleknejad}
\affiliation{Theoretical Physics Department, CERN, 1211 Geneva 23, Switzerland
}
\emailAdd{azadeh.maleknejad@cern.ch}

\abstract{This work introduces the chiral memory effect on the celestial sphere that measures the permanent change of electromagnetic fields by spin-dependent processes in bulk. Unlike the conventional memory effect based on the permanent soft shift in the gauge field itself, it is a permanent change in its spin angular momentum. The concept underlying the chiral memory (conventional memory) effect is optical spin torque (optical force) induction in bulk. Photons and EM radiation carry angular momentum, which is conserved without interactions. Chiral interactions with matter, medium, curvature, and theories with parity violation, i.e., axion-QED, transfers spin angular momentum to EM fields. In nature, such phenomena occur either on EM radiation (chiral memory) or in the vacuum of QED (vacuum chiral memory). It can be parametrized in terms of the photon's topological (axial) current at null infinity. To elude the gauge ambiguity of the topological current, we use the transverse gauge and show it is the physical part of the current suggested by its cohomology structure. 
 }

\begin{document}
\maketitle

\date{\today}

\section{Introduction}

\input{Intro}

\input{Setup}



\section{Discussion}\label{Sec:discussion}
\input{discussion}

\section*{\small Acknowledgment}

I am grateful to Eiichiro Komatsu for stimulating discussions on chiral radiation and to Sasha Zhiboedov for insightful discussions on memory effects. I would like to thank Tim Cohen, Mina Himwich, and Atul Sharma for helpful discussions. In memory of Zhina (Mahsa) Amini, the symbol and heroine of \textit{Woman + Life = Freedom}.

\appendix

\section{Details of Photons in Axion-QED}\label{Cal}
\input{details-axion.tex}

\bibliographystyle{JHEP}
\bibliography{references}

\end{document}

%% file: Intro.tex
The chirality of electromagnetic fields is an essential feature in many scientific areas, ranging from high-energy physics and cosmology \cite{Komatsu:2022nvu} to condensed matter, optics \cite{Mun20}, and quantum information \cite{Lodahl17}. Despite the chiral symmetry of the free electrodynamics in flat space, spin-dependent interactions of the photon and its environment can violate this symmetry. Such chiral interactions are ubiquitous in nature, and among them are astrophysical optically active/magnetized plasma, cosmological chiral phase transitions, and rotating massive objects. Consider $U(1)$ electromagnetism in an asymptotically flat geometry with chiral symmetry in most spacetime but a finite region. If the chiral interaction in bulk can leave a permanent change detectable on the celestial sphere, what is the null operator at $ \scrip $ that measures this effect? Moreover, what is the physical concept and conservation law underlying this memory effect?  Our aim in this work is to answer the above questions.


The conventional gravitational/electromagnetic memory effect refers to the plasticity of the spacetime/gauge field due to gravitational (GW)/electromagnetic (EM) radiation. The characteristic feature of these effects is the permanent displacement of a test/charged particle after a burst of GW/EM radiation passes \cite{Zeldovich:1974gvh, Bieri:2013hqa}. Recently, memory effects have attended new interest due to their intimate connection with symmetry structures of the asymptotically flat spacetimes and soft graviton/photon theorems \cite{Strominger:2013lka, Strominger:2014pwa, He:2014cra, Pasterski:2015zua}. Several types of gravitational memory effects have been discovered, e.g., gravitational spin memory effect \cite{Pasterski:2015tva} and gyroscopic memory effect \cite{Seraj:2021rxd}. \footnote{The gravitational spin memory is the time delay between clockwise and counterclockwise orbits near $\scrip$ induced by radiative (angular component of) momentum flux \cite{Pasterski:2015tva}. This time delay is linear in gravitational wave fields, i.e., $\Delta P(u) = \oint_{\mathcal{C}} (D^z C_{zz}dz + D^\bz C_{\bz\bz}d\bz)$.} 
 The conventional electromagnetic memory effect is the kick induced by the optical force \cite{Bieri:2013hqa} where the passage of EM radiation generates a kick to the charged particle as $\Delta \bv = \frac{q}{m} \int \bE dt$. That can be formulated in terms of a soft shift in the gauge field at asymptotic null infinity and it is equivalent to the soft photon theorem \cite{He:2014cra, Pasterski:2015zua, Susskind:2015hpa}. For further developments in this topic see \cite{Lysov:2014csa, Himwich:2019dug, Miller:2021hty, Pasterski:2021raf}




 
\begin{figure}
 \centering
\includegraphics[width=13.3cm]{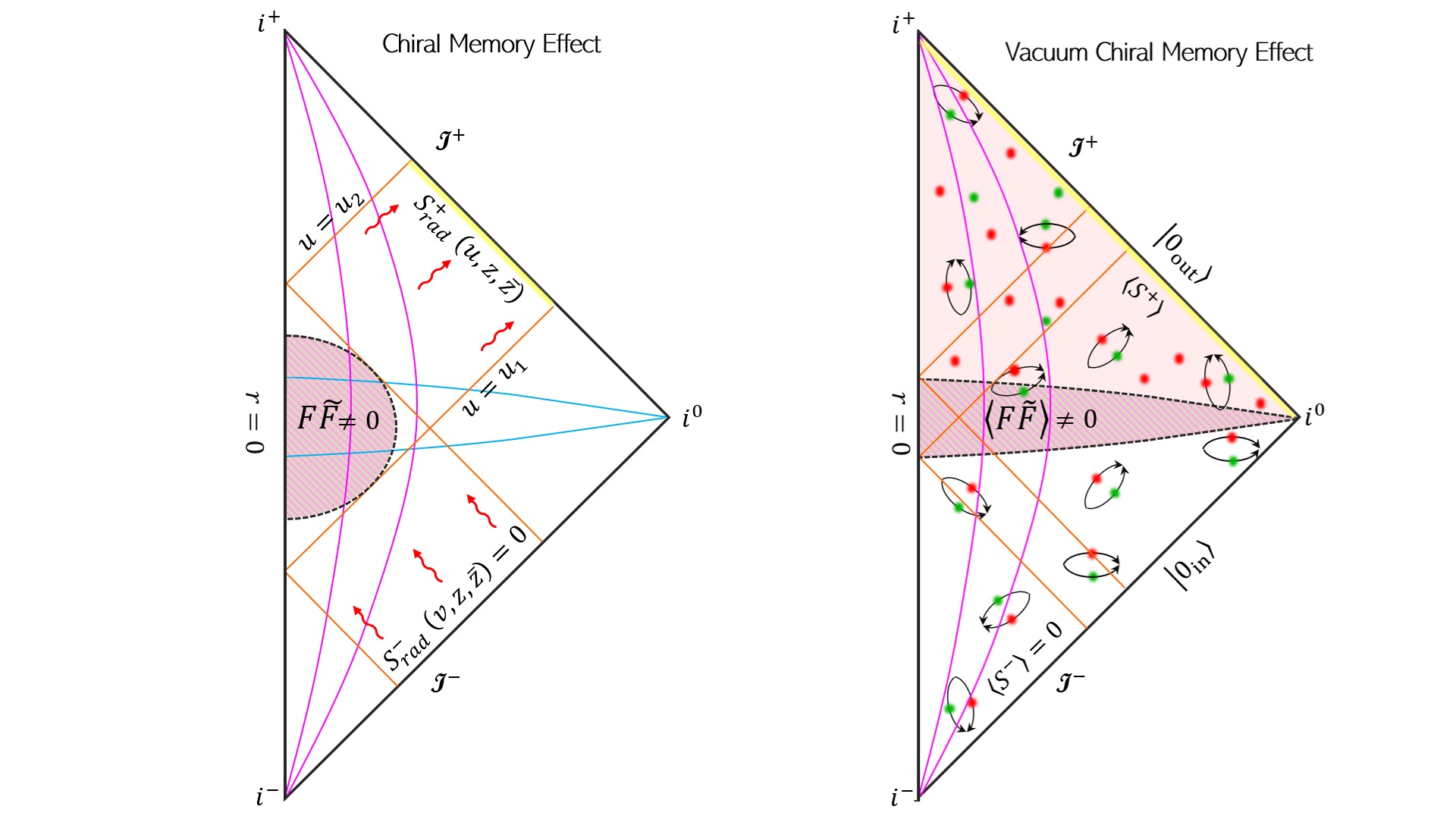}
\caption{The illustration shows the Penrose diagram of two types of chiral memory effects. Every point in the diagrams is a 2-sphere, $r=0$ excluded.  Left panel: A chiral source  (Shaded region) violates the conservation of helicity and spin angular momentum by optical torque induction for a period $u_1<u<u_2$ specified by its domain of dependence $\mathcal{D}^{+}$. However, helicity and spin are conserved before and after that. This short phase of parity violation generates a memory effect in the form of an EM spin flux  (helicity flux) at $\scrip$ denoted as $S^{+}_{rad}(u,z,\bz)$. Right panel: The process with vacuum quantum fluctuations. At $\scrim$ the system starts from an unpolarized vacuum state $\lvert 0_{in} \rangle$. Unequal amounts of left- and right-handed photons are generated during a chiral phase transition, i.e., $N_{R}\neq N_{L}$. Eventually, we get to another vacuum at $\scrip$. The new vacuum is chiral with a non-zero $\langle S^+\rangle$. We call this phenomenon the vacuum chiral memory effect to emphasize that it is a chiral effect imprinted in the vacuum state.
}\label{Penrose}
\end{figure}

\begin{figure}[h]
 \centering
\includegraphics[width=14cm]{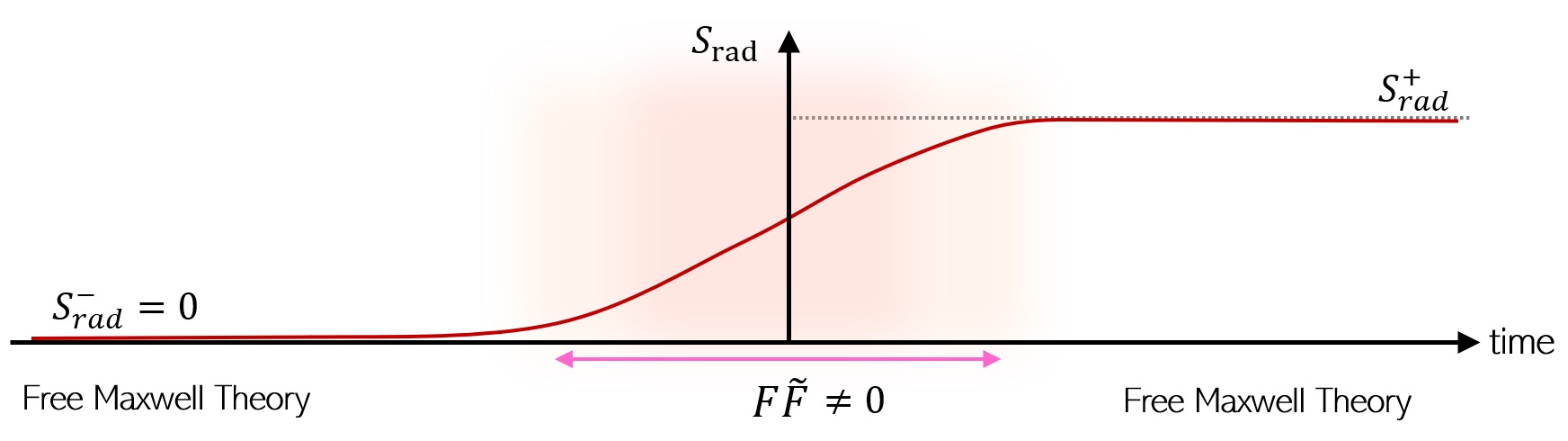}
\caption{At asymptotic past and future, we have free Maxwell theory. Local chiral interactions with matter/medium/curved space-time can induce spin torque and change the spin angular momentum of EM radiation. In the absence of interactions, the $S_{rad}$ remains conserved. Therefore, local chiral interactions lead to a permanent change in the spin angular momentum of EM radiation. }\label{conservation-S}
\end{figure}

In addition to linear momentum, EM fields also carry angular momentum, which is conserved in the absence of interactions. \footnote{In 1909, JH Poynting predicted light should carry angular momentum \cite{Poynting}, but concluded ``my present experience of light-forces does not give me much hope that the effect could be detected''. The effect has been measured 27 years later by Beth \cite{PhysRev.50.115}.  } Chiral interactions between EM fields and their environment transfer angular momentum between photons and matter in the form of optical torques. In this work, we show that this effect can be parameterized in terms of the photon topological current and has a finite value on the celestial sphere. In fact, it induces a permanent change to the EM spin angular momentum (SAM) and helicity, which can be observed at future null infinity $\scrip$ (See Fig. \ref{Penrose} and Fig. \ref{conservation-S}). The existence of such a memory effect relies on the fact that 
the free electromagnetic SAM and magnetic helicity are conserved in Minkowski space-time \cite{Cohen-Tannoudji:113864}.




This paper is structured as follows. Sec. \ref{Setup} discusses the setup of this work in terms of topological current of photon and its asymptotic expansion in asymptotically flat space-times. In Sec. \ref{EGA} we present an approach to elude the gauge ambiguity of topological current. In  Sec. \ref{Physical} we show the chiral observable is related to helicity flux and optical spin angular momentum. In
 Sec. \ref{memory-effect} we discuss that the chiral charge on $\scrip$ is the memory effect of optical spin torque transfer between EM fields and their environment.  The focus of Sec. \ref{Sec:light-ray} is on light-ray operators at $\scrip$ and putting our new observable into this context. In Sec. \ref{Sec:examples} we discuss two illustrative examples of the chiral memory effect. Finally we conclude in Sec. \ref{Sec:discussion}. Technical details of the computations of vacuum-vacuum transition in axion-QED is provided in Appendix \ref{Cal}.

\textbf{Notations and conventions: }  Throughout this work $j^{\mu}$ denotes Noether current, $j_c^{\mu} \equiv K^{\mu}$ is topological (axial) current, and $j_5^{\mu}$ is fermionic chiral current. The topological current in the transverse gauge is shown as $\mathcal{K}^{\mu}$. The $\bot$ and $\parallel$ indices denote the transverse and longitudinal contributions, respectively. We use bold symbols to denote 3-vectors. Here $\bJ$ is the angular momentum density, and $\bcJ$ denotes the total angular momentum. Besides, $\bS$ and $\bL$ are spin and orbital parts of the angular momentum. We abbreviate the complex angular coordinates as $\Theta^A=(z,\bz)$.  The vacuum permittivity is denoted as $\epsilon_0$, and $\varepsilon(z,\bz)$ is an arbitrary function of angles.


%% file: Setup.tex
\section{Chiral Asymmetry and Topological Current}\label{Setup}

 Consider electromagnetism in an asymptotically flat spacetime where the system has chiral symmetry almost everywhere but in a finite region $\mathcal{R}$. The nature of this chiral asymmetry  can be due to the spin dependent interactions of the EM fields in bulk. These effects can induce a non-zero $F_{\mu\nu}\tilde{F}^{\mu\nu}$ inside $\mathcal{R}$ and vanishes anywhere else, i.e.
\bea
F_{\mu\nu}\tilde{F}^{\mu\nu} \neq 0 \quad  \textmd{For} \quad x^{\mu} \in \mathcal{R},
\eea
where $\tilde{F}^{\mu\nu}= \frac12 \epsilon^{\mu\nu\lambda\sigma} F_{\lambda\sigma}$. Therefore, the left and right circular polarizations of (massless) photons have the same field equations everywhere but inside $\mathcal{R}$. We are interested in the observable effects of this chiral phenomenon at future null infinity $\scrip$ in asymptotically flat spacetimes. 
 It can be captured by an operator that measures the helicity of the field.     
 The related current operator is the photon topological (axial) current 
\bea
j_c^{\mu} = \epsilon^{\mu\nu\lambda\sigma} A_{\nu}F_{\lambda\sigma},
\eea
which satisfies the chiral anomaly equation
\bea\label{FtF}
\nabla_{\mu} j^{\mu}_c = F_{\mu\nu}\tilde{F}^{\mu\nu}.
\eea 
The subscript $c$ in $j^{\mu}_c$ denotes the chiral nature of the axial current compared with the vector nature of the Noether current $j^{\mu}$. We will discuss later that $j^0_c$ is the magnetic helicity density related to the imbalance between the left and right circular polarizations, and 3-vector $\bj_c$  is related to the helicity flux density. 

Here we introduce an arbitrary angular function $\varepsilon(\theta,\phi)$. That is inspired by the standard electrodynamics texts, in which the multipole expansion of the configurations is studied \cite{He:2014cra}. Integrating Eq. \eqref{FtF} over the entire spacetime convoluted by $\varepsilon(\theta,\phi)$, we find an integrated continuity equation as 
\bea\label{conti-Qc}
Q_{c}^+ = \int_{\mathcal{M}_4} \varepsilon(\theta,\phi) F\tilde F ~ - Q_{c}, 
\eea
where $Q_{c}^+ $ is the helicity flux at $\scrip$, $\mathcal{M}_4$ denotes the space-time manifold, and $Q_{c}$ is the contribution of the bulk's modes i.e.
\bea\label{Qc---}
Q_{c} = \int_0^{\infty} \sqrt{-g}d^3x ~ \varepsilon(\theta,\phi) ~ \big( j_c^{u} + \int  du \nabla_{A}j_c^{A} \big), 
\eea
where $u=t-r$ is the Bondi time and ``$A$'' index denotes the angular components, i.e. $\theta,\phi$. The $Q_{c}^+$ is our desired quantity given as
\bea\label{Qc+---}
Q_{c}^+ &=&  \int_{\mathcal{I}^+} \varepsilon(\theta,\phi) ~  j_c^{r}  \nonumber\\
 &=& -\lim_{r\rightarrow \infty} r^2 \int  du\sin\theta d\theta d\phi ~\varepsilon ~ j_{cu}.
\eea
In the second line we used the fact that in the asymptotically flat space-times for any 4-vectors we have $V^r=-V_u+\mO(\frac{1}{r^2})$ . The advantage of using the $\varepsilon(\theta,\phi)$ function in the definition of the chiral charge is that it can capture the angular distribution of the source. For instance, for a $F\tilde F$ with the structure of a dipole, $\int_{\mathcal{M}_4} F\tilde F=0$ but $\int_{\mathcal{M}_4} \cos\theta F\tilde F \neq0$.
 The noteworthy features of the above continuity equation are as follows. 
\begin{itemize}
\item{The first term of $Q_{c}$ with $\varepsilon(\theta,\phi)=1$ is the topological charge which is quantized for non-Abelian gauge fields in 4d. In electrodynamics, however, it is not quantized but still describes topological properties of the vector fields, e.g., twist, kink, knottedness, or linkage of the field lines \cite{1958PNAS44489W}. \footnote{The quantization of the topological charge is a characteristic property of the Yang–Mills theory in 4d and has its origin in the non-triviality of the mapping $\pi_3(SU(N))= \mathbb{Z}$. For the Abelian case, however, topologically non-trivial gauge transformations in 3d space do not exist since $\pi_3(U(1))=0$.}
}
 \item{The $Q^+_{c}$ term is a surface term usually neglected in the literature. After asymptotic expansion of the field at $r\rightarrow \infty$, we will show in Sec. \ref{AE} that $Q_{c}^+ $ can be non-zero and finite .}
\item{The permanent changes to the EM field due to interactions are stored on the asymptotic form of the gauge field near $\scrip$. $Q_{c}^+ $ is independent of the details of the source and is entirely specified by the asymptotic form of $A_{\mu}$. Nevertheless, it carries valuable information about the nature of the source to $\scrip$.}
\item{Both $Q_{c}$ and $Q_{c}^+$ are defined in terms of the axial current, which is not gauge invariant. To define a physical observable, one needs to elude the gauge ambiguity, which we will do in Sec. \ref{EGA}. }
\item{The $Q_{c}^+$ is the convoluted helicity flux at $\scrip$ (see Eq. \eqref{Eq-helicity-flux}) and related to the spin angular momentum of EM fields (see Sec. \ref{EM-Spin-Sub}). }
\item{ The physical concept underlying $Q^+_{c}$ is the permanent change to the polarization of EM radiation due to spin torque transfers in bulk (see Sec. \ref{Sec-Spin-Torque})
.}
\end{itemize}

\subsection{Asymptotic Expansion}\label{AE}

Now we turn to the case of asymptotically flat space-time with a particular interest in the physics near $\scrip$. The Minkowski space-time in retarded null coordinates is
\bea
ds^2 = -du^2 - 2dudr + 2r^2\gamma_{z\bz}dzd\bz,
\eea
where $z$ and $\bz$ are the complex spherical coordinate, and $\gamma_{z\bz} = \frac{2}{(1+z\bz)^2}$. Next, we need to fix the gauge at $\scrim$. Adopting the notations of \cite{Strominger:2013lka}, we choose the radiation gauge 
\bea\label{radiation-gauge}
A_u\rvert_{_{\scrim}}=0, \quad r^2\nabla^{\mu} A_{\mu}\rvert_{_{\scrim}}= 2(-r^2\p_u A_r + D^A A_{A})\rvert_{_{\scrim}}=0,
\eea
where $A$ index denotes angular coordinates $\Theta^A=(z,\bz)$, and $D^A$ is the covariant derivative with respect to $\gamma_{z\bz}$.
 Near the $\scrip$, the gauge field has the fall-off conditions \cite{Strominger:2013lka} \footnote{ Near the $\scrip$, the electromagnetic energy density and the electric force should satisfy the fall-off conditions $T_{uu} = \mO(r^{-2})$ and $F_{ur} =\mO(r^{-2})$.} 
\bea\label{fall-off}
A_\mu dx^{\mu} = \frac1r \mA_u(u,z,\bz) du + \frac{1}{r^2} \mA_r(u,z,\bz)dr +  A_{A}(u,z,\bz)d\Theta^A + \dots,
\eea 
where $\mA_{\mu}=\mA_{\mu}(u,z,\bz)$ represents the $r$-independent fields, and $\dots$ denote subleading terms in $r$. The gauge field takes different values at given directions specified by $\hat{\br}$ at $\scrim$ and $\scrip$ but obeys an antipodal matching condition as
\bea\label{matching--}
\lim_{r\rightarrow \infty} r^2 F_{ru}(\hat{\br})\vert_{\scrip_-} = \lim_{r\rightarrow \infty} r^2 F_{rv}(-\hat{\br})\vert_{\scrim_+},
\eea
where $\scrip_-$ is the past of $\scrip$ and $\scrim_+$ is the future of $\scrim$. 

The asymptotic form of the topological current at future null infinity is 
\bea\label{S+}
r^2j^c_u = \lim_{r\rightarrow \infty} 2\gamma^{z\bz}\big(\mA_{z}\p_{u} \mA_{\bz} - \mA_{\bz}\p_{u} \mA_{z}\big),
\eea
where $J^c_u= -J_c^r+\mO(\frac{1}{r^2})$ and the right hand side is finite and r-independent. Taking the limit $r\rightarrow \infty$, the dominant part of axial current is
\bea
\lim_{r\rightarrow \infty} r^2j_c^{\mu} \simeq (0,-r^2j^c_u,0,0), 
\eea
where the next to leading order terms are suppresed as $\mO(\frac{1}{r^2})$. 

\section{Eluding Gauge Ambiguity}\label{EGA}

We introduced $Q_c$ and $Q_c^+$ in term of $j^{\mu}_c$ in Eq.s \eqref{Qc---} and \eqref{Qc+---}. Despite $\nabla_{\mu} j_c^{\mu}$ is gauge invariant and a physical quantity, the $j_c^{\mu}$ itself is explicitly defined in terms of $A_{\mu}$ and is not.
 To construct a physical observable from $j_c^{\mu}$, we need first to elude the gauge ambiguity, which is the aim of this part. \footnote{I am grateful to Sasha Zhiboedov for insisting on this question.}

Under a gauge transformation \bea\label{gauge-trans}
A_{\mu} \mapsto A_{\mu} +\p_{\mu}\Lambda,
\eea
the chiral anomaly transforms as
\bea
\nabla_{\mu}j_c^{\mu}\mapsto \nabla_{\mu}j_c^{\mu} + 2 \nabla_{\mu}(\p_{\nu}\Lambda \tilde{F}^{\mu\nu}).
\eea
which imposing Bianchi identity, 
it proves $\nabla_{\mu}j_c^{\mu}$ is gauge invariant. In more abstract terms based on differential 1-forms, under a gauge transformation, the axial current transforms as $j_c \mapsto j_c' = j_c + d\tilde{\Lambda}$. Therefore, their difference is an exact form ($j_c$ and $j_c'$ are cohomologus) and $dj_c=dj_c'$.

Let us decompose the gauge field  as
\bea
A_{\mu} = A_{\mu \bot} + A_{\mu \parallel},
\eea
where the $\bot$ and $\parallel$ subscriptions denote the transverse and longitudinal contributions, respectively. By definition, we have \bea
A_{\mu \parallel} \equiv \p_{\mu} \lambda \quad \an \nabla^{\mu} A_{\mu\bot}=0,
\eea
where $\lambda$ is a scalar field.  Under a gauge transformation as Eq. \eqref{gauge-trans}, $A_{\mu\bot}$ is gauge invariant but $A_{\mu \parallel}$ transforms accordingly, i.e. 
\bea
A_{\mu\bot} \mapsto A_{\mu\bot}, \quad \qquad  A_{\mu\parallel} \mapsto A_{\mu \parallel} + \p_{\mu} \Lambda.
\eea
For the sake of convenience, we change the notation for the axial current as
\bea
K^{\mu} \equiv j^{\mu}_c.
\eea
We can decompose the axial current into two parts as
\bea
K^{\mu} = K^{\mu}_{\bot} + K^{\mu}_{\parallel}, 
\eea
where $ K^{\mu}_{\bot}$ is gauge invariant part \footnote{Note that the physical part of $A_{\mu\parallel}$ contributes to $F_{\mu\nu}$ (as $\bE_{\parallel}$) and hence to $ K^{\mu}_{\bot}$.}
\bea
 K^{\mu}_{\bot} = \epsilon^{\mu\nu\lambda\sigma}  F_{\lambda\sigma} A_{\nu \bot},
\eea
and $K^{\mu}_{\parallel}$ is the gauge dependent part 
\bea
 K^{\mu}_{\parallel} &=& \epsilon^{\mu\nu\lambda\sigma}  F_{\lambda\sigma} A_{\nu\parallel} \nonumber\\
 &=& \nabla_{\nu} \big(\epsilon^{\mu\nu\lambda\sigma} \lambda F_{\lambda\sigma}\big).
\eea
The above implies that the gauge-dependent part of the axial current is conserved. It is only the gauge-independent part that can be anomalous 
\bea\label{dKbot}
dK_{\bot} &=& F\tilde{F}, \\
 dK_{\parallel} &=&0.
\eea
 This is due to the cohomology structure of $K_{\mu}$ and related to the fact that $K=K_{\mu}dx^{\mu}$ and $K'=K'_{\mu}dx^{\mu}$ are cohomologous if they have the same $ K_{\mu\bot}$, i.e., $K'-K=dw$ where $dw$ is an exact 1-form. 
This was first pointed out in \cite{PhysRevA.86.013845} and is widely used in the optics literature. In the above we present the proof in more abstract form based on differential geometry. 

To summarize, although $j_c^{\mu} \equiv K^{\mu}$ is not gauge invariant, only its transverse (gauge invariant) part contributes to the anomaly equation. Since in this work, we are interested in the effect of spin torque, which induces a non-zero $F\tilde{F}$ in bulk, the $K^{\mu}_{\bot}$ has all the information we need. 
It shows that the physical part of the axial current can be unambiguously defined in the Lorenz gauge. This is the approach we adopted throughout this work. From now on we will use $K^{\mu}_{\bot}$ instead of $j_c^{\mu}$, and for shorthand we define
\bea\label{mK}
\mathcal{K}^{\mu} \equiv K^{\mu}_{\bot}= \epsilon^{\mu\nu\lambda\sigma} A_{\nu\bot} F_{\lambda\sigma}.
\eea
Hence the physical $Q_c$ and $Q^+_c$ are given in terms of $\mathcal{K}^{\mu}$.

\section{Helicity and Optical Angular Momentum}\label{Physical}
 
The integrands of $Q_{c}$ and $Q_{c}^+$ have interesting physical interpretations. Equation \eqref{dKbot} is a gauge invariant anomaly equation that can be written as a local helicity continuity equation as 
\bea
\p_t h + \bd.\bh  = 2 \bE.\bB,
\eea
where $\bE$ and $\bB$ are the electric and magnetic fields, $h$ is the magnetic helicity density \cite{Galaverni:2020xrq} 
\bea
h \equiv \frac12 \sqrt{-g} \mK^0 = \bA_{\bot}.\bB,
\eea
and $\bh$ is the helicity flux density
\bea\label{Eq-helicity-flux}
\bh \equiv \frac12 \sqrt{-g} \bmK^i =  \bA_{\bot} \times \bE.
\eea
Generally, the helicity of a given smooth vector field $X$ is defined as $\bX.(\bd\times \bX)$, and is related to the vorticity or twist of angular momentum of the field. It is a key quantity to measure the topological properties of the vector fields, e.g., twist, kink, knottedness, or linkage of the field lines  \cite{1999GMS...111..301B, app9050828}.
Magnetic helicity is a key quantity in plasma physics and optical fields to characterize the topology of magnetic lines \cite{1958PNAS44489W}. 

We can write $Q_c$ as
\bea
Q_{c} = 2 \int  d^3x ~\varepsilon ~ \bA_{\bot}.\bB +  2 \int  d^4x ~\varepsilon ~ \nabla_A(\bA_{\bot}\times\bE)^A, 
\eea
and physical $Q_{c}^+$ is
\bea\label{Q5-helicity}
Q_{c}^+ = 2\lim_{r\rightarrow \infty} r^2 \int  dud\Omega ~\varepsilon ~ \hat{\br}.(\bA_{\bot}\times\bE).
\eea
Upon setting $\varepsilon(\theta,\phi)=1$, $Q_c^+$ is the helicity flux at $\scrip$ and the first term of $Q_c$ is the total magnetic helicity.

\subsection{Spin Angular Momentum}\label{EM-Spin-Sub}

It turns out that $Q_c^+$ has a more physically intuitive form in terms of the spin angular momentum (SAM) of electromagnetic fields. Since the photon is massless, only the SAM component in the direction of propagation is physically meaningful, i.e., its helicity. In the following, we present helicity flux density in terms of the optical SAM flux. 

 The total angular momentum of the electromagnetic field $\bcJ$ is \cite{Cohen-Tannoudji:113864, book}
\bea\label{J-coul-rad}
{\bcJ}  \equiv \int  \bJ dx^3 = \bcJ_{Coul} + \bcJ_{rad},
\eea
where $\bJ=\br\times (\bE\times \bB)$, $\bcJ_{Coul}$ is the angular momentum of charged particles (Coulomb part), and $\bcJ_{rad}$ is the contribution of radiation.  \footnote{The separation \eqref{J-coul-rad} is based on using Helmholtz's theorem to decompose electric field into  longitudinal ($\bDel \times \bE_{\parallel}=0 $ ) and a transverse ($\bDel.\bE_{\bot}=0 $) parts
$\bcJ_{rad} = \int  \br \times (\bE_{\bot}\times \bB) dx^3$ and  
$\bcJ_{Coul} =\int \br \times (\bE_{\parallel}\times \bB) dx^3$.} Notice that $\bJ$ and $\bcJ$ denote the \textit{density} and \textit{total} quantities respectively. Using $\bDel.\bE_{\parallel} = \rho/\epsilon_0$ where  $\epsilon_0$ is the vacuum permittivity, one can write Coulomb part as the angular momentum of the charged particles
\bea
\bcJ_{Coul} = \sum_{i} \bJ_{i}.
\eea
Turning to the radiation part, $\bcJ_{rad} $ is related only to the transverse fields and conserved without interactions. For an electromagnetic field that falls off fast enough at $r\rightarrow \infty$, one can decompose the angular momentum into two separate parts as \cite{Cohen-Tannoudji:113864, book}
\bea
\bcJ_{rad} =   \epsilon_0 \int  E^i_{\bot}(\br\times \bDel)A^i_{\bot} dx^3 + \epsilon_0 \int  (\bE_{\bot}\times \bA_{\bot}) dx^3 ,
\eea
where the first term is associated with the orbital angular momentum density 
\bea
\bL_{rad}= E^i_{\bot}(\br\times \bDel)A^i_{\bot}.
\eea
and the second term is related to the spin angular momentum density 
\bea\label{Srad--}
\bS_{rad}= - \bA_{\bot} \times \bE_{\bot}.
\eea
The orbital angular momentum is related to helical phase fronts and optical vortices.
The spin angular momentum is the intrinsic part and independent of the definition of the origin of the coordinate system. It is associated with polarization.   Since only the transverse part of the vector potential enters the expressions for $\bL_{rad}$ and $\bS_{rad}$, the two contributions are both gauge independent. Combining Eq.s \eqref{Eq-helicity-flux} and \eqref{Srad--}, we see that SAM density of radiation and the helicity flux are related as
\bea
\bh = - \bS_{rad} - \sum_i \bS_{i},
\eea
where $\bS_i$ is the spin of spinning charged particles.


\section{Chiral Memory Effect}\label{memory-effect}

The conventional electromagnetic memory effect is the kick induced by the optical force \cite{Bieri:2013hqa} where the passage of EM radiation generates a kick to the charged particle as
\bea
\Delta \bv = \frac{q}{m} \int \bE dt.
\eea
The radial electric field corresponds to $F_{ru}$ and the above effect can be formulated in terms of $Q^+$ in Eq. \eqref{ASt-charge} \cite{He:2014cra, Pasterski:2015zua, Susskind:2015hpa}. 
In addition to linear momentum, EM fields also carry angular momentum, which is conserved in the absence of interactions. Chiral interactions with matter and curvature transfer angular momentum between photons and matter in the form of optical torques. That can induce permanent changes similar to the optical force.  This chiral memory effect can be formulated in terms of $Q_{c}^+$, which is studied in this work. Up to this point, we introduced $Q_{c}^+$ in terms of the topological (axial) current of the photon in Sec. \ref{Setup}. We consider free Maxwell theory at asymptotic regions $\scri^{\pm}$. Any possible chiral interactions with matter and curvature inside the bulk can generate a non-zero $F\tilde{F}$, which makes a permanent change to the topological current. 
Observers at $\scrip$ can detect this chiral memory effect by measuring the helicity flux, i.e., $Q_{c}^+$. In Sec. \ref{EM-Spin-Sub}, we showed that it is related to the flux of spin angular momentum at $\scrip$. That reveals that the chiral memory effect associated with $Q_c^+$ is related to the optical spin torque.

\subsection{Optical Spin Torque Induction}\label{Sec-Spin-Torque}

Interaction of EM radiation with spinning fields like electrons, optically active medium, and curved space-times with an angular momentum aspect like mergers can transfer spin angular momentum between EM radiation and the matter. The spin induction effect is the focus of this part. 

 One can construct a covariant version of the angular momentum density operator based on the 3rd-rank tensor 
\bea\label{M3}
M^{\mu\nu\lambda} = T^{\mu\nu} x^{\lambda} - T^{\mu\lambda} x^{\nu},
\eea
where the angular momentum density is $J_k = \epsilon_{ijk} M^{0ij}$. For a conserved energy-momentum tensor, this 3rd-rank tensor is also conserved. The total energy-momentum tensor is
\bea
T_{\mu\nu} = T_{\mu}^{em}+ T_{\mu\nu}^{m},
\eea
where  $T_{\mu}^{m}$ denotes the matter and medium part and $T_{\mu}^{em}$ is the symmetric electromagnetic energy-momentum tensor
\bea
T^{em}_{\mu\nu} = F_{\mu\lambda}F^{~\lambda}_{\nu} - \frac14 g_{\mu\nu} F_{\lambda\sigma} F^{\lambda\sigma}.
\eea
If the total energy-momentum tensor is conserved
\bea
\nabla_{\mu} M^{\mu\nu\lambda}=0,
\eea
then the electromagnetic field satisfies in the continuity equation \footnote{For standard Maxwell theory, we have $\nabla_{\mu} T^{\mu\nu}_{em}=j_{\mu}F^{\mu\nu}$, where the Lorentz force is $j_{\mu}F^{\mu\nu}=(\bj.\bE,j^0\bE+\bj\times \bB)$.}
\bea
\nabla_{\mu} M^{\mu\nu\lambda}_{em} 
= 2 j_{\mu} F^{\mu[\nu} x^{\lambda]} =- \nabla_{\mu} M^{\mu\nu\lambda}_{m} .
\eea
Notice that $J^{\mu}$ denotes the angular momentum density, and $j^{\mu}$ is the Noether current. The interactions induce a torque that transfers angular momentum between the electromagnetic fields and matter. In Minkowski space-time, we find the torque induction to the EM fields as
\bea
\frac{ d\bJ_{em}}{dt} = - \frac{d \bJ_{m}}{dt} - \epsilon_{ijk}\nabla_l M^{lij} \hat{\bx}^{k},
\eea
where $\hat{\bx}^{k}$ is the normal vector in the $x^k$ direction. In case the divergence of $M^{ijk}$ vanishes, we get the most simple torque induction relation as
\bea
\frac{ d\bJ_{em}}{dt} = - \frac{d \bJ_{m}}{dt} .
\eea

 As an example, one can consider charged Dirac fermions interacting with photons 
\bea
M^{\mu\nu\lambda}_m= -i \bar{\psi} \gamma^{\mu} D^{[\nu}\psi x^{\lambda]} + \bar{\psi} \gamma^{\mu} \Sigma^{\nu\lambda} \psi,
\eea
where $\psi$ is the Dirac 4-spinor, $\gamma_{\mu}$ are the Dirac matrices, $D_{\mu}$ is the spin connection, and $\Sigma^{\mu\nu}=\frac{i}{4}[\gamma^{\mu},\gamma^{\nu}]$. The angular momentum vector of the charged fermions in flat space vacuum is
\bea\label{bJ-e}
\bJ_{m} = \psi^{\dag} \br \times (-i\bDel) \psi + \frac12 \bar{\psi}\boldsymbol{\gamma}\gamma_5 \psi,
\eea
where $\gamma_5=i \gamma_0\gamma_1\gamma_2\gamma_3$ and we defined $\boldsymbol{\gamma}= \gamma^i \hat{\bx}_i$. The first term in Eq. \eqref{bJ-e} is the orbital angular momentum, and the second term is the spin angular momentum, which is the fermionic chiral current 
\bea
j^{\mu}_5 \equiv  \bar{\psi} \gamma^{\mu} \gamma_5 \psi.
\eea
 That leads to a spin torque as
\bea
\frac{d\bS_{m}}{dt} = \frac{d \bj_5}{dt}.
\eea
In nature, chiral plasma can have the above feature as well as a chiral magnetic effect. It is a macroscopic manifestation of a quantum effect, chiral anomaly, and is a 4d analog of the quantum hall effect in condensed matter. For more details and extensive reviews on this topic, see \cite{Kharzeev:2013ffa, Brandenburg:2021xmo}.
 Another example of the matter field is an optical medium with dielectric and magnetic responses. This case is more complicated. However, effectively one can write the time evolution of the spin angular momentum of the medium as \cite{Rivas:2002}
\bea
\frac{d\bS_{m}}{dt} = \bdd_m \times \bB + \bmu_m \times \bE,
\eea
where $\bdd_m$ is the electric dipole moment and $\bmu_m$ is magnetic moment of the medium.  Curved space-time with angular momentum aspects can also transfer spin to photons. The source-free Maxwell equations in curved space-time can formally be considered as equations in flat space-time in the presence of an optically active medium expressed through the space-time metric tensor components \cite{1957DoSSR.114...73S, Plebanski:1959ff}. Assuming the total angular momentum is conserved, the environment (matter/medium/space-time) spin torque induces an optical spin torque as
\bea
\frac{d\bS_{rad}}{dt} \sim  - \frac{d \bS_{m}}{dt} \quad \textmd{For } \quad t_1 < t<t_2,
\eea
where $(t_1,t_2)$ is the interaction period of EM fields with the environment. If the interactions are for a finite period, this spin and angular momentum transfer will be memorized by the electromagnetic fields.


Consider initial radiation with vanishing spin angular momentum, i.e. 
\bea
\bS_{rad}\lvert_{\scrim} =0.
\eea
which, in the absence of spin interactions, is conserved. The antipodal matching condition in Eq. \eqref{matching--} implies that in the past of $\scrip$ 
\bea
\bS_{rad}\lvert_{\scrip_-} =0.
\eea
 Spin interactions with matter and space-time can induce a spin torque to the EM fields during the interaction. As a result, at future null infinity, we have 
\bea
\bS_{rad}\lvert_{\scrip} \neq 0.
\eea
Therefore, chiral interactions leave a permanent change to the spin angular density of radiation at $\scrip$, i.e.
\bea
Q_{c}^+(z,\bz) = \int du ~ \hat{\br}.\bS_{rad} \neq 0.
\eea 
The left panel of Fig. \ref{Penrose} shows the Penrose diagram of such a process. One simple example of this effect is the interaction between EM radiation and cold magnetized intergalactic plasma, which we will discuss in Sec. \ref{Faraday}.

\subsection{Chiral Vacuum-Vacuum Transition}

 In this part, we consider a generic vacuum-vacuum transition generated by spin-dependent quantum mechanical interactions with QED vacuum and calculate the evolution of $\bS_{rad}$. We will show that it induces a permanent change in $\bS_{rad}$ and $Q^+_{c}$, which can be measured at $\scrip$.

\begin{figure}[h]
\centering
\includegraphics[width=14cm]{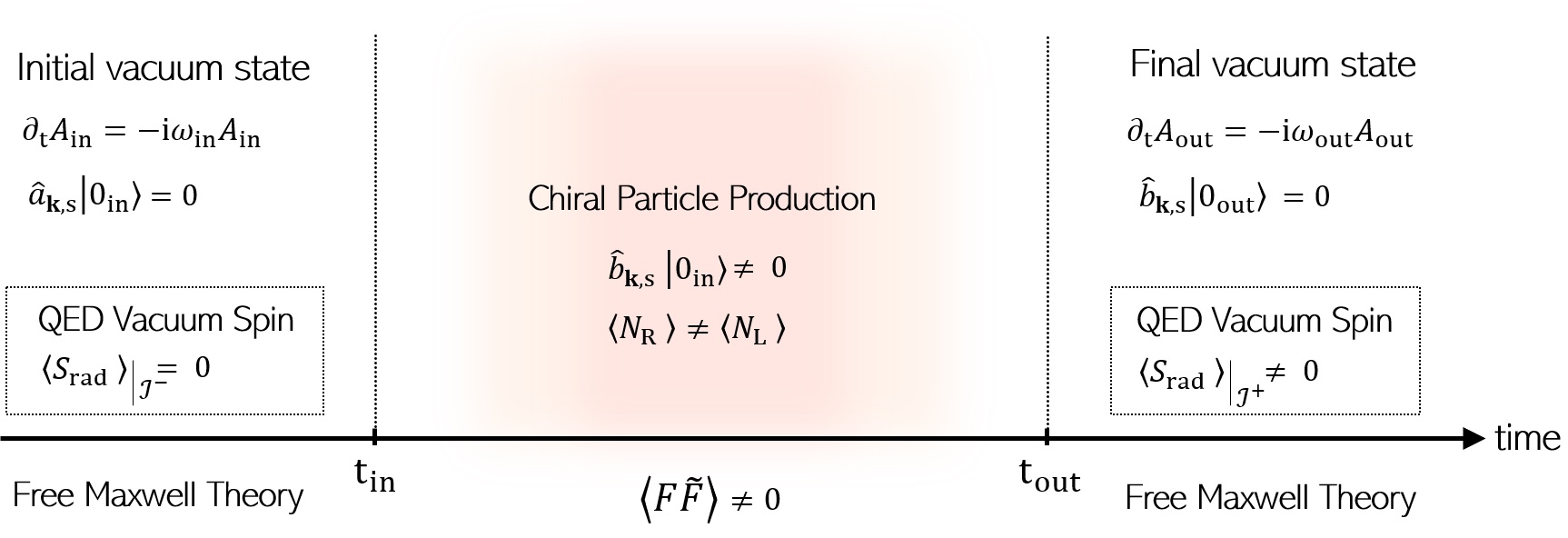}
\caption{Schematic figure of the chiral vacuum-vacuum transition process due to chiral interactions in bulk with chiral photon production and hence $\langle F\tilde{F} \rangle \neq 0$.}\label{Fig-ChiralVaca}
\end{figure}

Consider an initial (unpolarized) quantum vacuum at asymptotic past as
\bea
\lim_{t\rightarrow -\infty} \bA(t,\bx) =  \sum_{s=R,L }  \int \frac{d^3k}{(2\pi)^{\frac32}}\frac{1}{\sqrt{2k}}\bigg[  \hat{a}_{\bk,s} e^{ik.x}  \E_{s}(\hat{\bk}) +   \hat{a}^{\dag}_{\bk,s} e^{-ik.x} \E^*_{s}(\hat{\bk})\bigg],
\eea
where $\hat{a}^{\dag}_{\bk,s}$ and $\hat{a}_{\bk,s}$ are the creator and annihilation operators at $\mathcal{I}^-$ as
\bea
\hat{a}_{\bk,s} \vert 0_{\rm{in}} \rangle =0 \qquad \textmd{for all $\bk$},
\eea
and $\vert 0_{\rm{in}} \rangle$ is the vacuum state at  $\mathcal{I}^-$. Here $\hat{a}^{\dag}_{\bk,s}$ and $\hat{a}_{\bk,s}$ are the creator and annihilation operators
\bea
[\hat{a}_{\bk,s},\hat{a}^{\dag}_{\bk',s'}]= \delta_{ss'}\delta^3(\bk-\bk'),
\eea
 and $\E_{_{R,L}}(\hat{\bk})$ are the polarization states of the right- and left-handed photons propagating in $\hat{\bk}$ direction 
\bea
\E_{_{R,L}}(\hat{\bk}) = \frac{1}{\sqrt{2}}(\hat{\boldsymbol{\theta}}_{k} \pm i \hat{\boldsymbol{\phi}}_{k}),
\eea
where $(\hat{\bk}, \hat{\boldsymbol{\theta}}_{k}, \hat{\boldsymbol{\phi}}_{k})$ form a spherical basis vector set. 

The spin angular momentum density of the vacuum at $\scrim$ vanishes
\bea
\hat{\bS}\vert 0_{\rm{in}} \rangle =0.
\eea
The EM-matter interactions can produce particles and inject force and torque between the asymptotic past and future. Although we did not consider the Chern-Simon term in our EM theory, the chiral interactions that induce torque to EM fields generate a non-zero $\langle F\tilde{F} \rangle$. Eventually, these interactions are turned off, and the system gets to another vacuum at the asymptotic future
\bea\label{Aatscip}
\lim_{t\rightarrow +\infty} \bA(t,\bx) =  \sum_{s=R,L }  \int \frac{d^3k}{(2\pi)^{\frac32}}\bigg[ f_{s}(t,k) \hat{b}_{\bk,s} e^{ik.x} \E_{s}(\hat{\bk}) + f^*_s(t,k)  \hat{b}^{\dag}_{\bk,s} e^{-ik.x} \E^*_{s}(\hat{\bk})\bigg],
\eea
in which  $\hat{b}^{\dag}_{\bk,s}$ and $\hat{b}_{\bk,s}$ are the creator and annihilation operators at $\mathcal{I}^+$, i.e.
\bea
\hat{b}_{\bk,s} \vert 0_{\rm{out}} \rangle =0 \qquad \textmd{for all $\bk$},
\eea
where $\vert 0_{\rm{out}} \rangle$ is the vacuum state at  $\mathcal{I}^+$.  The positive frequency mode function at $\scrip$ is
 \bea
f_s(t,k) = \frac{1}{\sqrt{2\omega_{k,s}(t)}} e^{-i\int \omega_{k,s}(t')dt'},
\eea
where $\omega_{k,s}(t)$ is the effective frequency of the mode with momentum $\bk$ and helicity $s$, that satisfies the adiabatic conditions \cite{Parker:2009uva}. The $\hat{a}_{\bk,s}$ and $\hat{b}_{\bk,s}$ operators are related by Bogoliubov transformations 
\bea
\hat{b}_{\bk,s} = \alpha_{k,s}~  \hat{a}_{\bk,s} + \beta^*_{k,s} \hat{a}^{\dag}_{-\bk,s},
\eea
where $\alpha_{k,s} $ and $\beta_{k,s} $ are the Bogoliubov coefficients 
\bea
\lvert \alpha_{k,s} \rvert^2 - \lvert \beta_{k,s}  \rvert^2 = 1.
\eea
The expectation value of the photons with helicity state $s$ and momentum $\bk$ at $\mathcal{I}^+$ is 
\bea
\langle N_{\bk,s} \rangle =   \langle 0_{\rm{in}} \vert \hat{b}^{\dag}_{\bk,s} \hat{b}_{\bk,s} \vert 0_{\rm{in}}  \rangle = \vert \beta_{k,s} \vert^2.
\eea
Finally, The vacuum-vacuum transition can be computed as
\bea
\lvert \langle 0_{\rm{out}} \vert 0_{\rm{in}} \rangle \rvert^2 =  e^{-\iint d^3x dt \Upsilon_{\rm{vac}}} = \exp\big[ -\frac{1}{(2\pi)^3} \sum_{s=\pm} ~ \int d^3x \int d^3k \ln(1+ \lvert \beta_{k,s}\rvert^2) \big],
\eea
where $\Upsilon_{\rm{vac}}$ is the vacuum decay rate.  Using the large $r$ saddle-point approximation with fixed $u$, we can expand $A_z$ as \cite{Lysov:2014csa}
\bea
A_{z}(u,z,\bz) =  - \frac{i}{8\pi^2}\sum_{s=R,L}  \int_0^{\infty} \sqrt{2k} dk \bigg[  \hat{a}_{\bk,s} e^{-iku}  -  \hat{a}^{\dag}_{\bk,-s} e^{iku} \bigg]~ [\E_{s}(\hat{\bk})]_z.
\eea
 One can expand $A_{\bz}$ accordingly. That gives the spin flux operator at $\scrip$ as
\bea\label{S-Q}
\hat{\br}.\bS^+(u) = \hbar \int_0^{\infty} k^2dk \big[ \hat{N}_{\bk R}(u) - \hat{N}_{\bk L}(u)\big],
\eea
 where $\hat{N}_{\bk,s}$ is the number operator of photons in the $(\bk,s)$ modes, i.e. $\hat{N}_{\bk,s}\lvert_{\scrim}=\hat{a}^{\dag}_{\bk,s} \hat{a}_{\bk,s}$ and $\hat{N}_{\bk,s}\lvert_{\scrip}=\hat{b}^{\dag}_{\bk,s} \hat{b}_{\bk,s}$.  Inserting \eqref{Aatscip} in \eqref{S-Q}, we find the initial value of $\hat{\bS}$ as
 \bea
 \langle \hat{\br}.\hat{\bS} \rangle \vert_{\scrim}=0,
 \eea
while its value in the asymptotic future is
\bea
\langle \hat{\br}.\hat{\bS}(u) \rangle\vert_{\scrip}= \hbar \int k^2dk ~ \hat{k} \big(  \lvert \beta_{kR}(u) \rvert^2 - \lvert \beta_{kL}(u) \rvert^2 \big).
 \eea 
Fig. \ref{conservation-S} and Fig. \ref{Fig-ChiralVaca} schematically illustrate the chiral vacuum-vacuum transition phenomenon. The right panel of Fig. \ref{Penrose} shows the Penrose diagram of such a process. One simple example of this phenomenon can be seen in axion-QED, which we will discuss in Sec. \ref{axion-QED-sec}.

\section{Operators on Celestial Sphere}\label{Sec:light-ray}

 To contextualize our discussion, let us compare our result with the standard analysis of asymptotic symmetries in terms of light-ray operators.  Light-ray operators are non-local operators that naturally emerge from integrating Einstein equations at null infinity along light sheet time, i.e. $u$. We start this section with a quick overview of light-ray operators in the literature.

\subsection{Light-ray Operators} 

Charge operators are the integral of a 4-current $j_{\mu}$ defined on the co-dimension one sub-manifold $\Sigma^{\mu}$. The future charge operator is defined as \cite{He:2014cra} 
\bea\label{ASt-charge}
Q^+ = \lim_{r\rightarrow +\infty} \int_{\scrip_{-}} r^2 d\Omega ~ \varepsilon F_{ru},
\eea
which can be decomposed as  \footnote{
The relevant constraint equation at $\scrip$ is
$r^2 \p_u F_{ru} + D^A F_{uA} + e^2r^2  j_u =0$. Therefore, the RHS of Eq. \eqref{decom-Q} has another surface integral over $\scrip_+$. Reference \cite{He:2014cra} set this term to zero by assuming that the electric field will vanish at $\scrip_+$.}
\bea\label{decom-Q}
Q^+ = Q^+_S + Q^+_H.
\eea
Here $Q^+_H$ is the hard charge defined as
\bea
Q^+_H = e^2\int_{\scrip} r^2d\Omega du ~\varepsilon ~ j_u,
\eea
where $j_{\mu}$ denotes the Noether current and $Q^+_S$ is a soft term linear in $F_{\mu\nu}$ as
 \bea
Q^+_S = \int_{\scrip} dud\Omega ~\varepsilon ~\big[ D^z F_{uz} + D^{\bz} F_{u\bz}\big] = \lim_{r\rightarrow \infty}\int d\Omega ~\varepsilon ~\big[ D^z \Delta\mA_{z} + D^{\bz} \Delta \mA_{\bz}\big],
\eea
where $\Delta \mA_{z} \equiv \mA_{z}\vert_{\scrip_+} - \mA_{z}\vert_{\scrip_-} $. 

Now let us consider gauge field configurations that satisfy the fall-off conditions \eqref{fall-off} and behave like a gauge transformation at the future and past of the $\scrip$, i.e.
\bea
A_{\mu}(u,\bx) \rightarrow \p_{\mu} \Lambda_{\pm} \quad \textmd{as} \quad   u  \rightarrow \pm \infty,
\eea
where $\Lambda_{+}$ and $\Lambda_{-}$ are two large gauge transformations at $\scrip_+$ and $\scrip_-$ respectively. That implies these gauges must be $u$-independent $\Lambda_{\pm}=\Lambda_{\pm}(z,\bz)$,
where in general, these two can deviate from each other, i.e., $\Delta \Lambda= \Lambda_{+} - \Lambda_{-}\neq 0$. 
Computing the soft charge by the above condition, one finds
\bea
Q^+_S =  2 \int dzd\bz ~\p_{z}\p_{\bz}\varepsilon ~ \Delta\Lambda.
\eea
The soft mode is the difference of the $A_{z}$ between the past and future of $\scrip$ convoluted with $\p_{z}\p_{\bz}\varepsilon $. For more details, see \cite{He:2014cra, Pasterski:2015zua, Susskind:2015hpa}. For a clear-cut and concise review of the recent developments in this topic, see \cite{Strominger:2017zoo}. Another operator linear on gauge field at $\scrip$ is the dipole moment \cite{Lysov:2014csa}
\bea
-e^2 \boldsymbol{\mathcal{P}}+2i\pi \boldsymbol{\mu} = \int d\Omega r^2 F_{zr} \p_{\bz} \hat{\bx}.
\eea
The 3rd rand operator $M_{\mu\nu\lambda}$ in \eqref{M3} can be used to define the charge assoated with the Lorentz group as \cite{Cordova:2018ygx}
\bea
\bcJ_{\mu\nu} = \int d\Sigma^{\lambda} T_{\lambda[\nu]} x_{\lambda},
\eea
and the total angular momentum is 
\bea
\bcJ_{k} = \epsilon^{ijk} \int d\Sigma^0 T_{0i} x_{j}.
\eea

Now we turn to light-ray operators. One can construct the following light-ray operators  based on the energy-momentum tensor $T_{\mu\nu}$ \cite{Cordova:2018ygx, Gonzo:2020xza, Hu:2022txx} are
\bea 
\mathcal{E}(z,\bz) &=& \lim_{r\rightarrow \infty}\int du ~ r^2T_{uu} ,\\
\mathcal{K}(z,\bz) &=& \lim_{r\rightarrow \infty}\int du ~ r^2 uT_{uu},\\
\mathcal{N}_{A}(z,\bz) &=& \lim_{r\rightarrow \infty}\int du ~ r^2T_{uA},
\eea
and the following based on the Noether current
\bea
\mathcal{Q}(z,\bz) = \lim_{r\rightarrow \infty}\int du ~ r^2j_u.
\eea
The above surface densities are energy density, boost energy density, momentum flux density, and charge density. The charge density $\mathcal{Q}(z,\bz)$ is related to the hard charge $Q_H^+$ as
\bea
Q_H^+ = e^2 \int d\Omega ~\varepsilon(z,\bz)~ \mathcal{Q}(z,\bz).
\eea
  Notice that $\mathcal{N}_{A}$ is literately the angular component of momentum flux which differs from EM angular momentum. The explicit form of $T_{uA}$ for EM in the radiation gauge is
\bea
r^2 T_{uA} = 2 \big( 2\gamma^{BC}\p_u A_{B} \p_{[A} A_{C]}  + r^2 \p_uA_{A} \p_u A_r\big),
\eea
where the $A,B,C$ indices denote $(z,\bz)$. For more details about light-ray operators and their algebra see \cite{Cordova:2018ygx, Gonzo:2020xza, Hu:2022txx}.

\subsection{Optical Spin Flux Operator}\label{EM-Spin-Operator-Sub}


In addition to the above operators based on the Noether current and the energy-momentum tensor, one can construct an operator in terms of the topological current $j^{\mu}_c$ (in transverse gauge as Eq. \eqref{mK})
\bea\label{Q5epsilon-}
Q_{c}^+(z,\bz) &=& - \lim_{r\rightarrow \infty} \int  du ~ r^2\mathcal{K}_u \\
&=&  ~ ~ 2 \lim_{r\rightarrow \infty} \int  du  ~ \gamma^{z\bz}\big( \mA_{\bz} \p_{u} \mA_z - \mA_{z} \p_{u} \mA_{\bz} \big),
\eea
which is the helicity flux density on the celestial sphere. It turns out that $Q_{c}^+$ has yet a more intuitive physical interpretation related to the spin angular momentum of radiation, $\bS_{rad}$. Using \eqref{Srad--}, we can rewrite $Q^+_{c}$ as
\bea\label{Qc++}
Q^+_{c}(z,\bz) =  -2 \lim_{r\rightarrow \infty}  \int_{-\infty}^{+\infty} r^2 du ~ \hat{\br}.\bS_{rad}(u,z,\bz).
\eea
Therefore, the new light-ray operator we defined based on topological current is the SAM flux of radiation at $\scrip$. The reason that the orbital angular momentum of radiation does not contribute to $Q^+_{c}$ is that $\hat{\br}.\bL_{rad}=0$.

The noteworthy features of $Q^+_{c}$ are as follows. i) It is entirely specified by the asymptotic form of the gauge field
on $\scrip$, ii) it is quadratic in $A_{z}$ and $A_{\bz}$, iii) it is parity odd, but iv) it is not soft, i.e., it can not be written as a large gauge transformation at $\scrip$. The features (ii)-(iv) differs between $Q^+_S$ and $Q^+_{c}$. Besides, $Q^+_{H}$ and $Q^+_c$ are different since the former is related to the vector current $j_u$, and the former is related to the axial current $j^c_u$.

\subsection*{Quantum Spin flux Operator at large r}

Up to this point, our discussion was based on classical electrodynamics in asymptotically flat space-times. Let us get closer to the real world and turn on quantum fluctuations. In optics, it is well-established that a beam of light can carry both spin and orbital angular momenta parallel to its propagation axis \cite{Cohen-Tannoudji:113864, book}. However, there has been confusion if they are separately meaningful since they are not true quantum operators for angular momenta. On the other hand, for a light beam
in the paraxial limit \footnote{In Geometric optics, a ray is called paraxial if the wave vectors of the field fall within a narrow cone with a small opening angle. }, the separation $\bJ_{rad}=\bL_{rad} +\bS_{rad}$ is well understood \cite{PhysRevA.45.8185}.

Upon a rotation about $\hat{\bn}$ direction by an angle $\phi$, a vector field $\bX$ transforms as
\bea
\bX(t,\bx) \rightarrow \bR_{\hat{\bn}}(\phi) \bX\big(t, \bR^{-1}_{\hat{\bn}}(\phi) \bx \big).
\eea
For an infinitesimal rotation with angle $\delta\phi \ll 1$, the vector field changes as
\bea
\delta \bX(t,\bx)  =  \big( \hat{\bn}.(\br\times \bDel) \bX(t,\bx) + \hat{\hat{\bn}}\times \bX(t,\bx)\big) ~ \delta\phi,
\eea
where $\hat{\bL}=-i \br \times \bDel$ is the orbital angular momentum operator, and the second term, related to the field's vectorial nature, is the spin angular momentum operator $\hat{\bS}$. Therefore, $\hat{\br}.\hat{\bS}$ is the generator of an infinitesimal rotation in the $\hat{r}$ direction. It is part of the Lorentz group in the representation of the target operator $\mathcal{O}$. \footnote{An analogous analysis can be done in Fourier space for transverse photon fields, $\bq.\bA(t, \bq)=0$. A rotation in $\bk$ direction in reciprocal space transforms the field as $\delta \bA(t,\bq)  = \delta\phi \big( \hat{\bk}.(\hat{\bq}\times \bDel_{\bq}) \bA(t,\bq) + \hat{\bk}\times \bA(t,\bq)\big)$. A generic infinitesimal rotation defined by $\hat{\bL}$ or $\hat{\bS}$ operators break down of the transversality condition, and only rotations defined by $\hat{\bJ}$
respects it \cite{Keller}. Nevertheless, $\hat{\bq}.\hat{\bS}$ also respects the transversality condition.} For instance, on a vector operator $\mO$ it acts as
\bea
[\hat{\br}.\hat{\bS}, \mO]  = i\hat{\br} \times \mO.
\eea 
Note that the $\hat{\br}.\hat{\bL}=0$ holds as an operator equation. Turing on quantum fluctuations, the
$A_{\mu}$ field can be well approximated by the free field near asymptotic regions $\scri^{\pm}$ 
\bea
\bA(t,\bx) =  \sum_{s=R,L }  \int \frac{d^3k}{(2\pi)^{\frac32}}\frac{1}{\sqrt{2k}}\bigg[  \hat{a}_{\bk,s} e^{ik.x} \E_{s}(\hat{\bk}) +   \hat{a}^{\dag}_{\bk,s} e^{-ik.x} \E^*_{s}(\hat{\bk})\bigg].
\eea
From Eq. \eqref{S-Q}, we find that $\hat{\br}.\bS^+_{rad}(u)$ is a measure of the imbalance between left and right-handed polarizations of photons.

\section{Two Examples}\label{Sec:examples}

Chiral phenomena that can leave chiral memories on celestial sphere are ubiquitous. One can mention optically active astrophysical media, gravitational Faraday effect analogical to the magneto-optical Faraday effect \cite{1957DoSSR.114...73S, Plebanski:1959ff}, axion-QED, etc. This section aims to illustrate further the physical consequences of the chiral memory effect with two more familiar examples that can be solved analytically. We first consider the Faraday rotation due to magnetized plasma, which is a common astrophysical and cosmological effect. As the second example, we consider the chiral vacuum effect induced by a phase transition in axion-QED.

\begin{figure}[h]
\centering
\includegraphics[width=14cm]{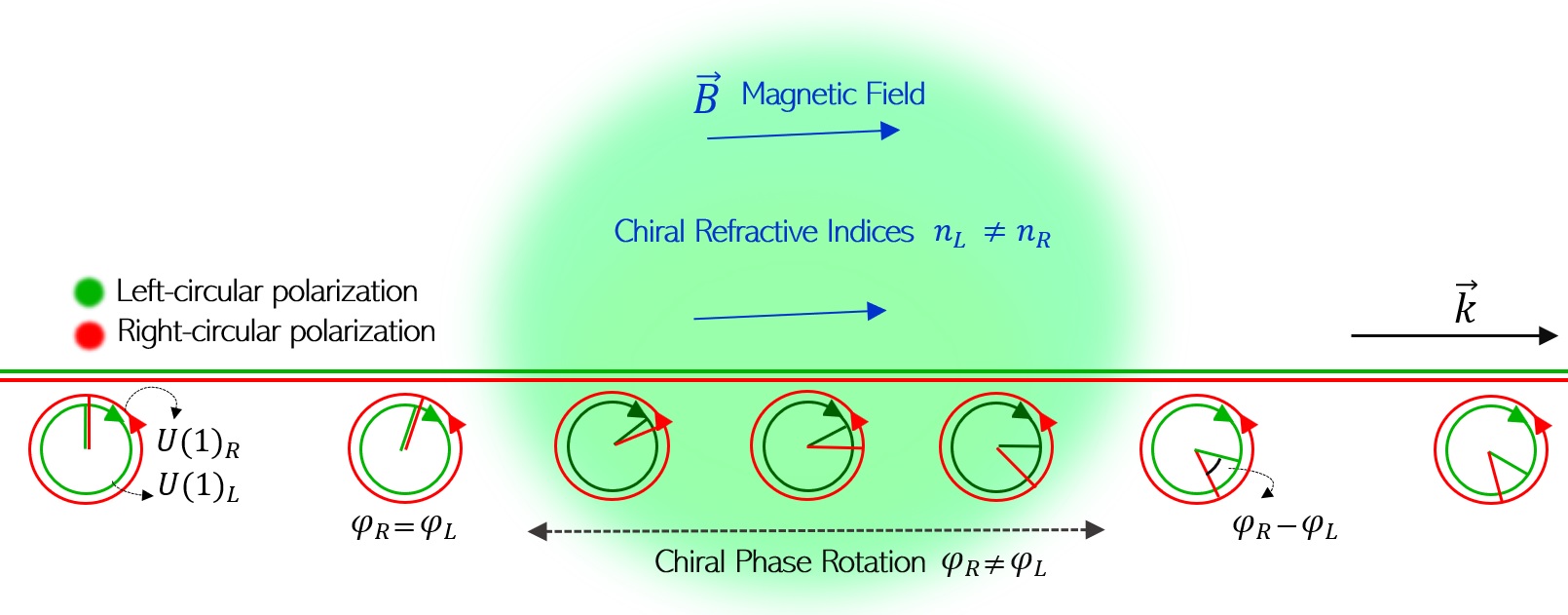}
\caption{ Faraday rotation. The magnetic field breaks parity and creates different dispersion relations for left- and right-hand circular waves propagating parallel to its direction, i.e., chiral refractive indices $\rm{n}_L\neq \rm{n}_R$. The left and right-handed polarizations of light propagating through such medium take two different phases, $\Delta\varphi=\varphi_R-\varphi_L \neq 0$. }
\label{Fig-Faraday}
\end{figure}

\subsection{Magnetized Plasma (Faraday Rotation)}\label{Faraday}

As the first example, we consider the Faraday rotation 
, i.e., rotation of the polarization plane of photons in a magnetic field $\bB$, due to the interaction of the spin (helicity) of the photons with the electrons in the medium. \footnote{Unlike the polarization rotation due to optical activity, Faraday rotation is non-reciprocal. The Faraday rotation of light with two opposite propagation directions through a magneto-optic material does not cancel but doubles. That is an essential component in optical telecommunications and other laser applications.}  We want to explore the chiral memory effect by Faraday rotation.  In astrophysical systems and cosmology, Faraday rotation naturally arises when light propagates through magnetized cold plasma, e.g., Earth's ionosphere, intergalactic regions, etc. Measuring Faraday rotation is a promising method to study the galactic and extragalactic magnetic fields. For a clear and concise review of Faraday rotation, see \cite{2021MNRAS.507.4968F}.

Consider a cold, magnetized plasma with an electron density $n_e$, and a magnetic field which, without loss of generality, we can consider in the $\hat{z}$ direction as 
\bea\label{B0}
\bB_{0}(t,\bx) =   \begin{cases}
B_0(t,\bx) \hz, & \quad \textmd{For}  \quad t_1 < t <t_2, \quad r< R, \\
0 & \quad  \textmd{Otherwise},  \quad
\end{cases}
\eea
where $B_0(t)$ is a (slow varing) quasi-static magnetic field and $R$ is a rough radius of the magnetized plasma.
The magnetic field breaks parity (symmetry under space reversal) and creates different dispersion relations for left- and right-hand circular waves propagating parallel to the direction of the magnetic field. In the $\omega \gg \omega_B, \omega_p$ limit, the dispersion relation of photons with polarization states  $\E_{R,L}(\hz)=\frac{1}{\sqrt2}(\hx\pm i \hy)$ are
\bea
\omega^2_{R,L}(k) \simeq \omega_0^2 \big[ 1 - \frac{\omega_p^2}{2\omega_0^2} ( 1 \mp \frac{\omega_B}{\omega_0}) \big] ,
\eea
where $\omega^2_p=\frac{n_e e^2}{\varepsilon_0 m_e}$ and $\omega_B=\frac{eB_0}{m_e}$ are the plasma and cyclotron frequencies respectively, $n_e$ is the number density of electrons, and $\omega_0=ck$ is the frequency of light in vacuum. The magnetized plasma has different refractive indices (phase velocities) for left- and right-handed $n_{R,L}=v_{R,L}/c$, i.e., anisotropic plasma.  That leads to a phase difference between the left- and right-hand circularly polarized components of light inside the medium as
\bea
\Delta\varphi(t,\bk) &\approx & \frac{1}{c}\int_0^d (\omega_R - \omega_L) dl\\
&=& \frac{e^3/ m_e^2}{2c^3\varepsilon_0k^2} \int_0^d dl~ n_e \bB(t,\bx).\hat{\bk} ,
\eea
which is related to the integral of the magnetic field parallel to its propagation direction, i.e., $B_{\parallel}=\bB(t,\bx).\hat{k}$. See Figure \ref{Fig-Faraday} for an illustration.

Consider an outgoing initial unpolarized radiation at asymptotic past as 
\bea\label{Avac}
\bA(t,\bx) = A_0 {\rm{Re}}\bigg[\big(\E_R(\hat{\bk})+\E_L(\hat{\bk}) \big) e^{i(\bk.\bx-kt)}\bigg] .
\eea
The initial radiation has zero spin density
\bea
\bS^-_{rad}(v,z,\bz) = \bA_{\bot}\times \bE_{\bot}\lvert_{_{\scrim}}=0. 
\eea
The light rays that get to $\scrip$ without interacting with the magnetized plasma will keep the form \eqref{Avac} with vanishing spin density. Now consider a magnetized plasma with magnetic field \eqref{B0}. The intersection of the future domain of dependence of the magnetized plasma region and $\scrip$ specifies a finite sector at $\scrip$ with $u_1<u<u_2$ that will receive the Faraday rotated light rays. \footnote{The future domain of dependence of a region $\Sigma_B$, denoted as $D^+(\Sigma_B)$, specifies all the points $p$ in spacetime that are connected to $\Sigma_B$ by a past inextendable causal curve. }  After passing through the magnetized medium and getting back to the vacuum, the radiation takes the form
\bea
\bA_{_\bB}(t,\bx) = A_0 {\rm{Re}}\bigg[\big(e^{i\varphi_R(t)}\E_R(\hat{\bk})+e^{i\varphi_L(t)}\E_L(\hat{\bk}) \big)  e^{i(\bk.\bx-kt)}\bigg].
\eea
That generates a non-zero spin density at $\scrip$ between $u_1<u<u_2$ as
\bea 
\bS^+_{rad}(u,z,\bz) = \p_u(\varphi_R-\varphi_L) A_0^2 \big[ 1 + \cos(2ku-2kr\sin^2\theta/2+\varphi_R+\varphi_L)\big] \hat{\bk},
\eea
where $\theta$ is the angle between $\hat{\br}$ and $\hat{\bk}$ (See Fig. \ref{conservation-S} and also the left panel of Fig. \ref{Penrose}).

\subsection{Axion QED (Chiral Vacuum)}\label{axion-QED-sec}

As for the second example, consider an axion QED theory. Unlike previous examples, in this case, we explicitly have a $F\tilde{F}$ term in the theory which is coupled to a dynamical field. We want to explore the vacuum transition and the change of spin angular momentum of radiation in the presence of a phase transition.
The electromagnetic part of the axion-QED action is 
\bea\label{axion-QED}
\mathcal{L} = -\frac{1}{4} F_{\mu\nu}F^{\mu\nu} - \frac{ \chi}{4\Lambda} F_{\mu\nu}\tilde{F}^{\mu\nu}.
\eea
where $\Lambda$ is a constant related to the decay constant of the axion. Now we consider a phase transition that breaks the Peccei-Quinn symmetry and gives a vacuum expectation value to the axion. At a time $t_1$ the symmetry is spontaneously broken
\bea
\langle \chi(t,\bx) \rangle \neq 0 \quad \textmd{for $t>t_1$, }
\eea
and after a time $t_2$ ($t_1<t_2$), the axion reaches a constant background value, i.e.
\bea
\langle \chi(t,\bx) \rangle = \chi_0 \quad \textmd{for $t>t_2$. } 
\eea
For the sake of simplicity and concreteness, we assume this phase transition can be approximately described as
\bea
\frac{ \langle \dot\chi \rangle }{\Lambda} =  \begin{cases}
 0 & t< t_1 \quad \textmd{and} \quad t> t_2\\
 \xi & \qquad t_1 < t < t_2 \\
  \end{cases}
\eea
where $\xi$ is a slow-varying function of time. Without loss of generality, we assume $\xi$ is always positive.

We are interested in the form of the photon fields at $\mathcal{I}^+$ when the electromagnetic fields start in a (quantum) vacuum at $\scrim$. In Coulomb gauge and Fourier space, we can expand the fields as
\bea
\bA(t,\bx) =  \sum_{s=R,L }  \int \frac{dk^3}{(2\pi)^{\frac32}}\bigg[  A_{s, k}(t)  \hat{a}_{\bk,s} e^{i\bk.\bx} \E_{s}(\hat{\bk})+  A^{*}_{s, k}(t) \hat{a}^{\dag}_{\bk,s} e^{-i\bk.\bx} \E^*_{s}(\hat{\bk}) \bigg],
\eea
where $\E_{R,L}(\hat{\bk}) = \frac{1}{\sqrt{2}}(\hat{\boldsymbol{\theta}}_{k} \pm i \hat{\boldsymbol{\phi}}_{k})$ are the plus/minus helicity state polarization vectors. The modified photon field before and after the phase transition ($t\geq t_1$ and $t \leq t_2$) is 
\bea
\p_t^2 A_{R,L} + k^2 A_{R,L}=0,
\eea
and during the phase transition ($t_1 \leq t \leq t_2$) is
\bea
\p_t^2 A_{R,L} + \omega^2_{R,L}(k)  A_{R,L} =0,
\eea
where $\omega^2_{R,L}(\bk)$ is the effective frequency squared 
\bea
\omega^2_{R,L}(k)  =k^2 \mp \frac{ \langle \dot{\chi} \rangle }{\Lambda}k.
\eea
  Since we assumed that $\xi>0$, the $\omega_{L}(k)$ is always positive definite, i.e. $\omega_{L}(k) = k \sqrt{1+\frac{\xi}{k}}$.  However, the value of $\omega_{R}(k)$ can be either positive, zero, or imaginary based on their momentum as 
\bea
\omega_{R}(k) = \begin{cases}
k\sqrt{(1 - \frac{\xi}{k})} & \textmd{for $k\geq \xi$},\\
i k \sqrt{(\frac{\xi}{k}-1)} & \textmd{for $k < \xi$}.
\end{cases}
\eea
For the opposite case with negative $\xi$, the $\omega_R$ is always positive, and it is the $\omega_L$ that can be negative.

\begin{figure}[h]
 \centering
\includegraphics[width=7cm]{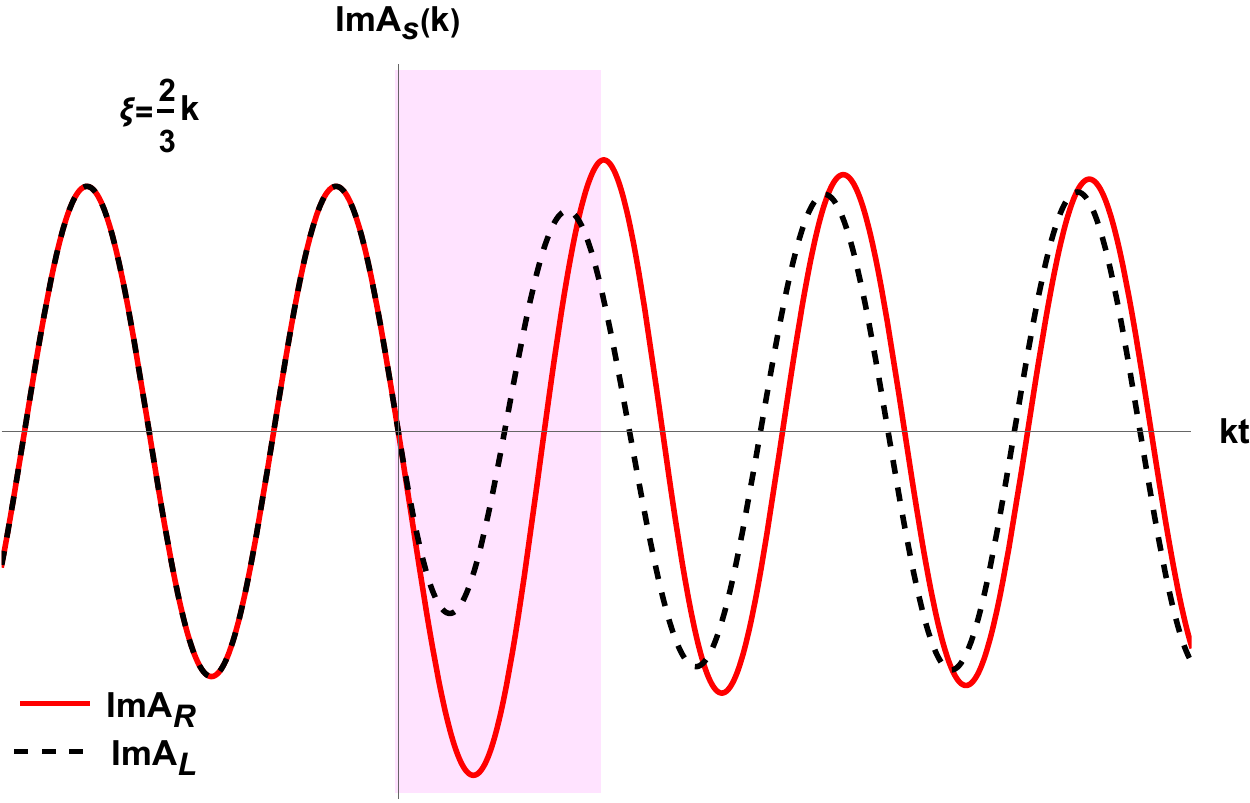}
\includegraphics[width=7cm]{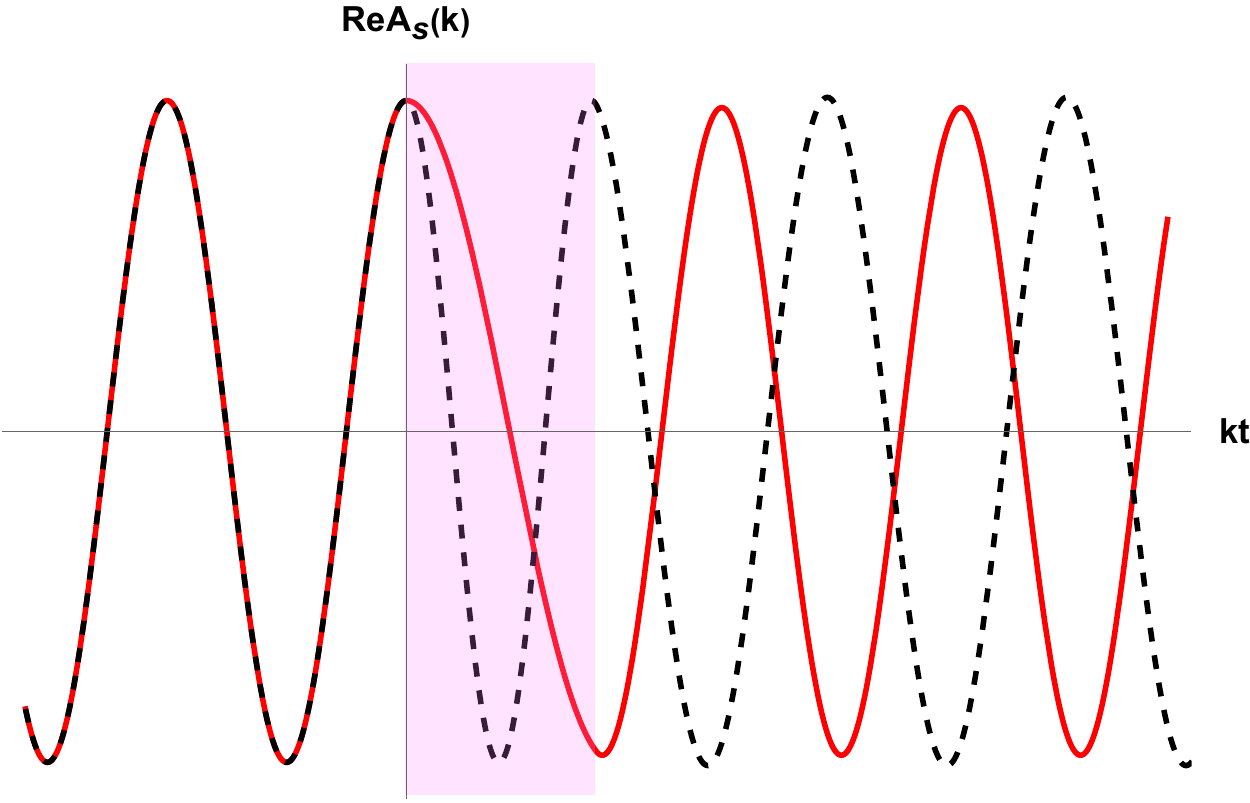}\\
\includegraphics[width=7cm]{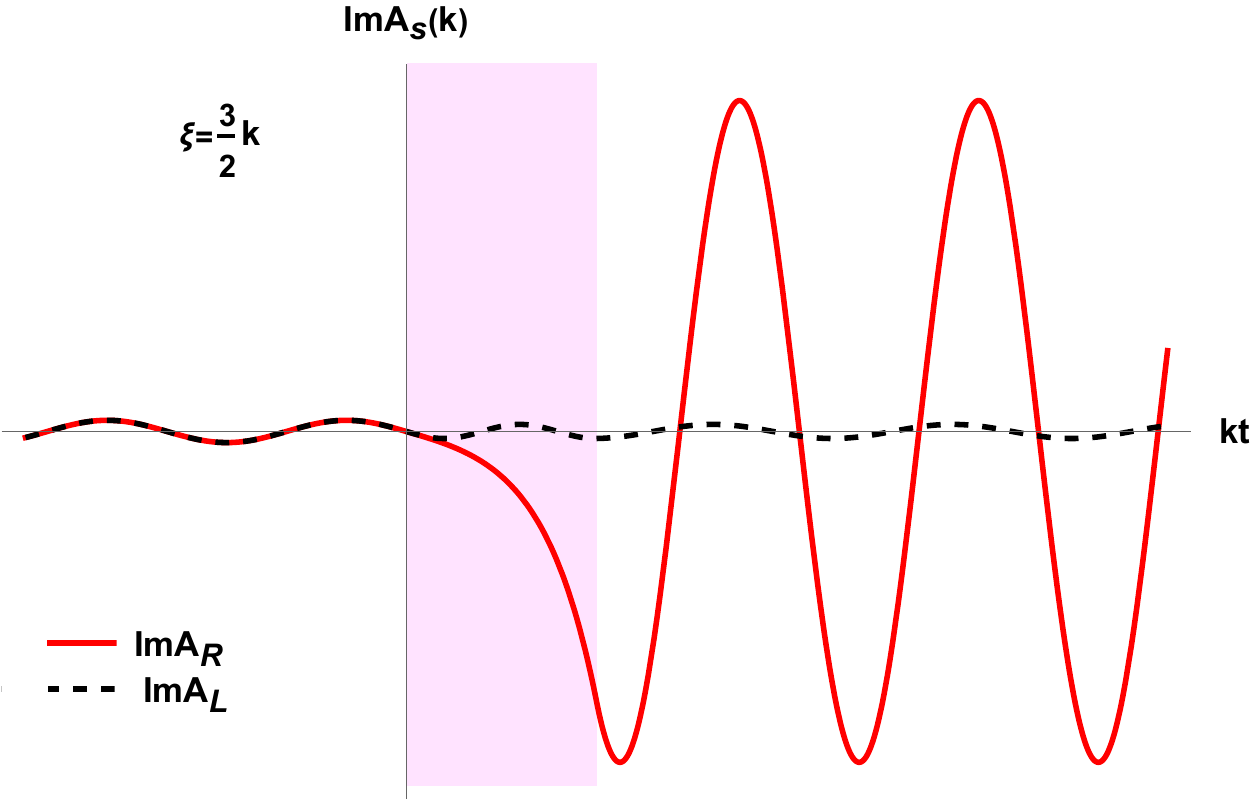}
\includegraphics[width=7cm]{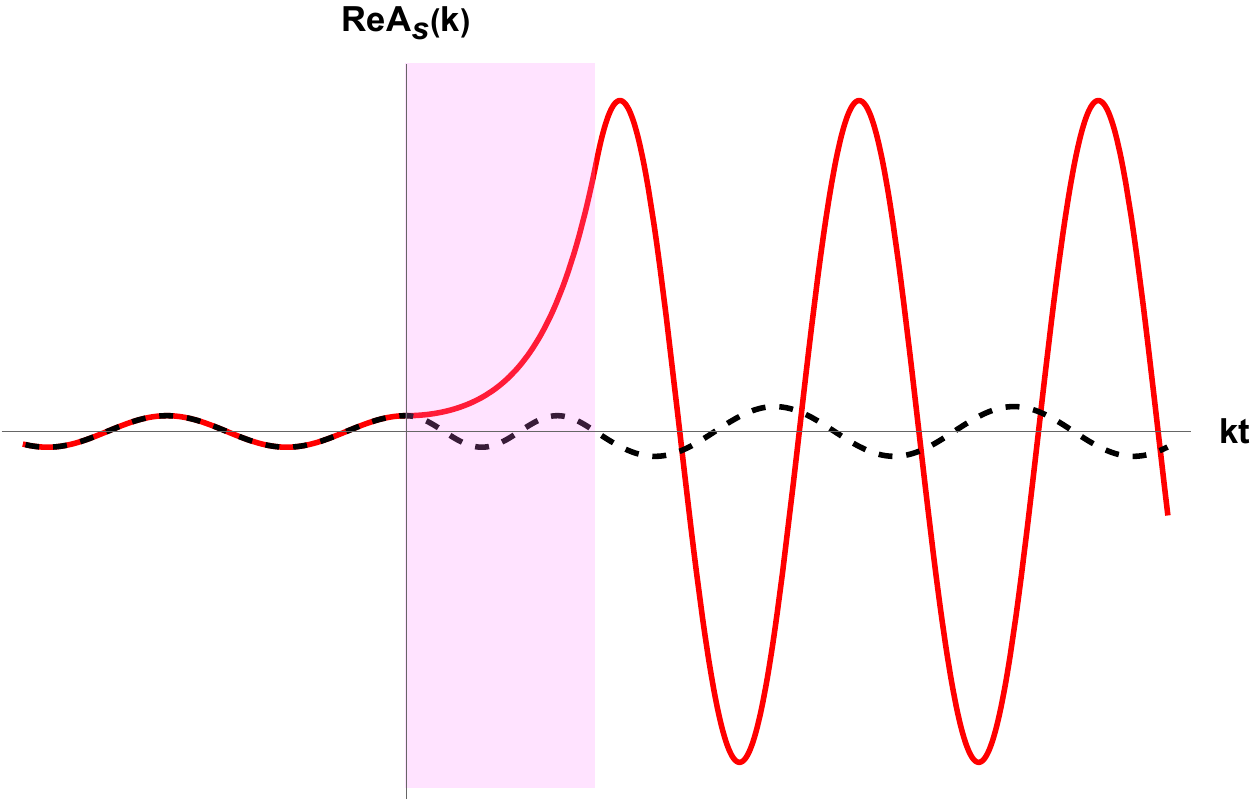}
\caption{Time evolution of mode functions for a given $k$ with $\xi=\frac23k$ (top) and $\xi=\frac32k$ (bottom). The horizontal axis is $kt$. The red (solid) line and the black (dashed) line show $A_{R}$ and $A_{L}$, respectively. The shaded area shows the phase transition interval $t_1<t<t_2$.}
\label{Plot-Apm-}
\end{figure}

We divide the system into three regions $I$, $II$, and $III$, corresponding to before, during, and after the phase transition, respectively.  Setting the initial conditions in the region $I$ with quantum vacuum fluctuation, we solve the mode functions in regions $II$ and $III$ with the matching condition at the boundary and the conservation of Wronskian. The details of calculations are presented in Appendix \ref{Cal}, and here we only report the final results.  Setting the initial value of $A_{R,L}$ as unpolarized vacuum fluctuations (i.e. an outgoing wave at $\mathcal{I}^-$) we find the mode functions in $t\leq t_1$ as
\bea
A^{I}_{s,k}(t) = \frac{1}{\sqrt{2k}} e^{-ikt}.
\eea 
During the phase transition $t_1 \leq t \leq t_2$, each helicity state has a different frequency, and the mode functions take the form 
\bea
A^{II}_{s}(t,k) = \big[ c^{s}_1 e^{-i\omega_{s}(k) t} + c^{s}_2 e^{i\omega_{s}(k)t}  \big],
\eea
where  $c^{R,L}_{1,2} $ are given by the matching condition $A^{I}_{R,L}(k,t_1) = A^{II}_{R,L}(k,t_1)$ and $\p_t A^{I}_{R,L}(k,t_1) = \p_t A^{II}_{R,L}(k,t_1)$. For the propagating modes, the latter corresponds to the conservation of Wronskian. See Eq.s \eqref{Cpm1}- \eqref{Cpm2} and \eqref{TCpm1}-\eqref{TCpm2} for the explicit form of $c^s_1$ and $c^s_2$. The solution in region III can be written as
\bea
A^{III}_{s}(t,k) = \big[ d^{s}_1(k) e^{-ik t} + d^{s}_2(k) e^{ikt}  \big],
\eea
where $d^s_1(k)$ and $d^s_2(k)$ coefficients are presented in Eq.s \eqref{ds1}-\eqref{ds2} and \eqref{ds1m}-\eqref{ds2m}. The time evolution of the mode functions $A_{R,L}(t,k)$ are presented in Fig. \ref{Plot-Apm-}.  As we see, the amplitude of $A_R$ with $k<\xi$ (bottom panels) gets amplified during the phase transition.

\subsubsection*{Particle Creation and vacuum-vacuum transition amplitude}

\begin{figure}[h]
 \begin{center}\label{Plot-Number}
\includegraphics[width=7cm]{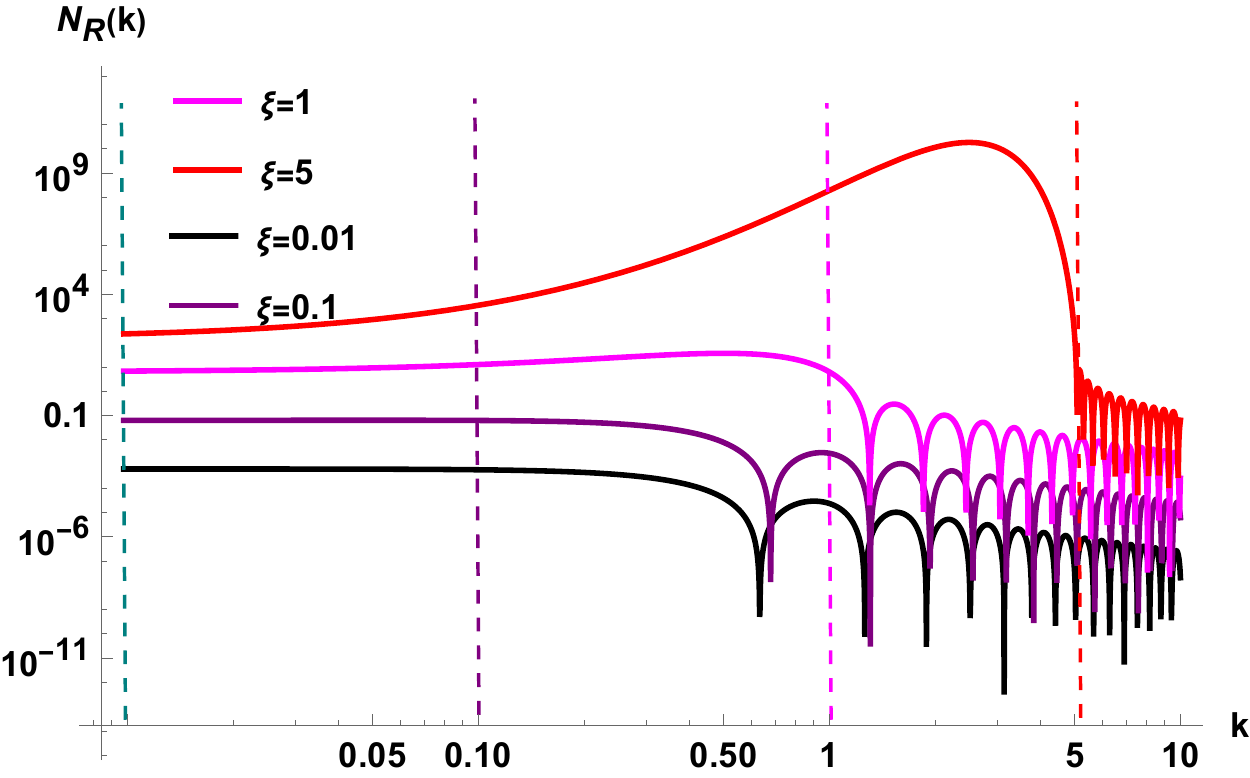}
\includegraphics[width=7cm]{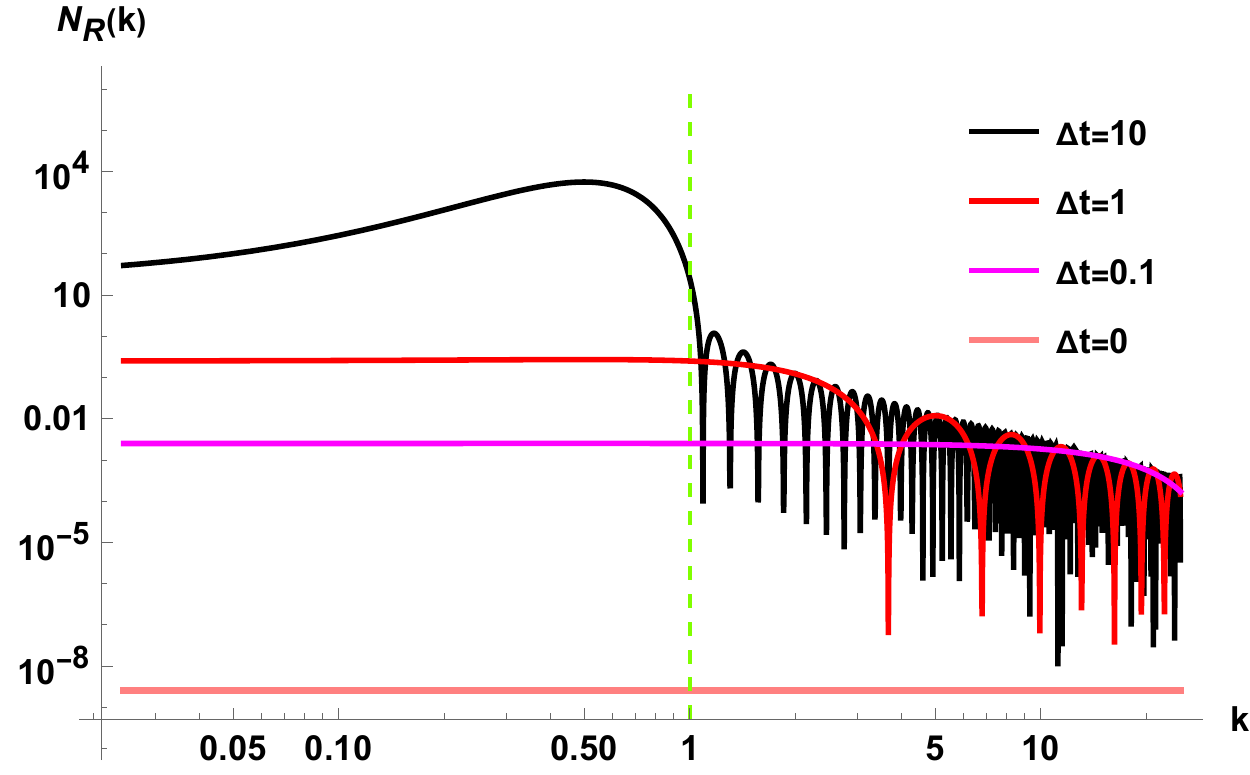}
\caption{The number density of right-handed photons $N_{\bk R}$ in terms of momentum $k$. The left and right panels show $N_{\bk R}$ in terms of different values of $\xi$ and $\Delta t$ respectively.}
\end{center}
\end{figure}

We have started from initial (unpolarized) QED vacuum fluctuations at $\mathcal{I}^-$. Interaction with a dynamical axion modifies the vacuum to the following form at $\mathcal{I}^+$
\bea
 \bA(t,\bx) = \lim_{r\rightarrow \infty} \sum_{s}  \int \frac{dk^3}{(2\pi)^{\frac32}}\bigg[  \hat{b}_{\bk,s} A^{III}_{s}(t,k)  e^{i \bk.\bx} \E_{s}(\bk) +   \hat{b}^{\dag}_{\bk,s} A^{III*}_{s}(t,k)  e^{-i \bk.\bx} \E^*_{s}(\bk)\bigg],~~~
\eea
in which $\hat{b}^{\dag}_{\bk,s}$ and $\hat{b}_{\bk,s}$ are the creator and annihilation operators at $\mathcal{I}^+$. Now using the Bogoliobov technique, we find the expectation value of the photons with helicity state $s$ and momentum $\bk$ at $\mathcal{I}^+$ as
\bea
\langle N_{\bk s} \rangle &= &  \langle 0_{\rm{in}} \vert \hat{b}^{\dag}_{\bk,s} \hat{b}_{\bk,s} \vert 0_{\rm{in}}  \rangle = \vert d_{k,s} \vert^2 \\
&=& \bigg( \frac{\xi~\Delta t }{2}\bigg)^2 \bigg(\frac{\sin \big[\omega_s(k)\Delta t \big]}{\omega_s(k)\Delta t }\bigg)^2,
\eea
where $\Delta t=t_2-t_2$. Notice that for plus helicity state with $k<\xi$, we have
\bea
\langle N_{\bk R} \rangle = \bigg( \frac{\xi~\Delta t }{2}\bigg)^2 \bigg(\frac{\sinh \big[\lvert \omega_R(k)\rvert\Delta t \big]}{\lvert \omega_R(k) \rvert\Delta t }\bigg)^2,
\eea
which can be very large. Fig \ref{Plot-Number} shows $N_{\bk R}$ for different values of $\xi$ and $\Delta t$. We illustrate this process in the right panel of Fig. \ref{Penrose}. 

 That induces a net chirality at $\scrip$ which can be described in terms of $\hat{\br}.\hat{\bS}(u)$ as 
\bea\label{Sepsi-}
\hat{\br}.\hat{\bS} &=& \hbar \int k^2dk  \big(  \hat{N}_{\bk R} - \hat{N}_{\bk L} \big)\nonumber\\
&\simeq & \hbar \bigg( \frac{\xi~\Delta t }{2}\bigg)^2 \int_0^{\xi} k^2dk ~\bigg(\frac{\sinh \big[\lvert \omega_R(k)\rvert\Delta t \big]}{\lvert \omega_R(k) \rvert\Delta t }\bigg)^2.
\eea
Since we set $\xi>0$, this spin induction effect is due to enhancement of $N_R$ and the final result is positive. In the opposite case with $\xi<0$,  the $N_L$ is enhanced and the spin flux is negative.

%% file: discussion.tex

We introduced a new observable on the celestial sphere in terms of the topological (axial) current of photon $j^{\mu}_c$, called $Q^+_{c}$ (Eq. \eqref{conti-Qc}). Our setup is free Maxwell theory at past and future null infinities of asymptotically flat spacetimes. Then possible chiral interactions inside the bulk can generate a non-zero $F\tilde{F}$, which can be seen as an anomaly equation for the topological current. More precisely, chiral interactions can create an imbalance between left and right-handed polarizations of photons and induce a helicity flux on $\scrip$ parametrized by $Q_{c}^+$. In Sec \ref{EGA}, we showed that writing the topological current in the transverse gauge fixes the gauge ambiguity since only the transverse part of the topological current can be anomalous.   The noteworthy features of $Q^+_{c}$ are as follows. It is entirely specified by the asymptotic form of the gauge field on $\scrip$, it is parity odd and quadratic in the gauge field. Therefore, unlike the Noether charge $Q^{+}$ based on $j^{\mu}$, $Q^+_c$ does not include a soft term and can not be written as a large gauge transformation on $\scrip$.


The $Q_c^+$ operator has an intuitive physical interpretation in terms of the spin angular momentum of electromagnetic fields $\bS_{rad}$ (Eq. \eqref{Qc++}). The SAM parameter is used to measure the degree of handedness of an optical field. Without any interactions with the environment,  $\bS_{rad}$ is conserved. However, chiral interactions can transfer spin angular momentum between EM fields and their environment. The concept underlying this phenomenon is the optical spin torque induction. It can be formulated as a continuity equation for the 3rd-rank tensor $M^{\mu\nu\lambda}$ in Eq. \eqref{M3}, which is the covariant version of the angular momentum density operator.



 Such chiral interactions are ubiquitous in nature, and among them are astrophysical optically active/magnetized plasma, interaction with spinning fields, rotating massive objects, and theories with parity violation. The chiral memory effect can occur either on EM radiation (chiral memory) or in the vacuum of QED (vacuum chiral memory). The former changes radiation polarization, while the latter induces vacuum chiral birefringence. Finally, we discussed two familiar examples of this effect that can be solved analytically. We considered the Faraday rotation due to magnetized plasma in Sec. \ref{Faraday} and chiral vacuum-vacuum transition by a phase transition in axion-QED in Sec. \ref{axion-QED-sec}.

%% file: details-axion.tex
This appendix is devoted to the details of the axion-QED calculations of Eq. \eqref{axion-QED} with a Peccei-Quinn phase transition as
\bea
\frac{ \langle \dot\chi \rangle }{\Lambda} =  \begin{cases}
 0 & t< t_1 \quad \textmd{Region I}\\
 \xi & t_1 < t < t_2 \quad \textmd{Region II}\\
  0 & t> t_2 \quad \textmd{Region III}\\
  \end{cases}
\eea
where $\xi$ is an almost constant. Without loss of generality, we assume that $\xi$ is always positive.

 In Coulomb gauge and Fourier space, we can expand the quantum field as
\bea
\bA(t,\bx) =  \sum_{s=R,L }  \int \frac{dk^3}{(2\pi)^{\frac32}}\bigg[  A_{s, k}(t)  \hat{a}_{\bk,s} e^{i\bk.\bx} \E_{s}(\bk) +  A^{*}_{s, k}(t) \hat{a}^{\dag}_{\bk,s} e^{-i\bk.\bx} \E^*_{s}(\bk)\bigg]~ ,
\eea
where $\E_{s}(\bk)$ is the plus/minus helicity state polarization vectors. We divide the system into three regions $I$, $II$, and $III$, corresponding to before, during, and after the phase transition, respectively. We solve the mode functions in each region and set the initial conditions in the region $I$ with quantum vacuum fluctuation and the initial conditions in regions $II$ and $III$ with the matching condition at the boundary and the conservation of Wronskian. \\
{\bf{Region $I$ ($t\leq t_1$):}} The field equation takes the following form
\bea
\p_t^2 A_{R,L} + k^2 A_{R,L}=0.
\eea
Imposing the vacuum fluctuations as the initial states at $\mathcal{I}^{-}$, gives the mode functions in region $I$ as
\bea
A_{R,L}(t,k) = \frac{1}{\sqrt{2k}} e^{-ikt}.
\eea\\
{\bf{Region $II$ ($t_1 \leq t\leq t_2$):}} The field equation during the phase transition is
\bea
&& \p_t^2 A_{R,L} + \omega^2_{R,L}(k)  A_{R,L} =0,
\eea
where $\omega^2_{R,L}(k)$ is the effective frequency squared for each helicity state given as
\bea
\omega^2_{R,L}(k)   = k^2(1 \mp \frac{\xi}{k}).
\eea
Since we assumed that $\xi>0$, the $\omega_{L}(k)$ is always positive definite i.e.
\bea
\omega_{L}(k) = k \sqrt{1+\frac{\xi}{k}}.
\eea
 However, the value of $\omega_{+}(k)$ can be either positive, zero, or imaginary based on their momentum as 
\bea
\omega_{R}(k) = \begin{cases}
k\sqrt{(1 - \frac{\xi}{k})} & \textmd{for $k\geq \xi$},\\
i k \sqrt{(\frac{\xi}{k}-1)} & \textmd{for $k < \xi$}.
\end{cases}
\eea
i) For $k > \xi:$ The most generic solution of the above equation is
\bea\label{As}
A_{s}(t,k) = \big[ c^{s}_1 e^{-i\omega_{s}(k) t} + c^{s}_2 e^{i\omega_{s}(k)t}  \big],
\eea
where  $c^{R,L}_{1,2} $ are specified by the matching conditions, i.e. $A^{I}_{R,L}(k,t_1) = A^{II}_{R,L}(k,t_1)$
\bea
\big[ c^{s}_1 e^{-i\omega_{s}(k)t_1} + c^{s}_2 e^{i \omega_{s}(k) t_1}  \big] = \frac{1}{\sqrt{2k}} e^{-ikt_1},
\eea
and $\p_t A^{I}_{R,L}(k,t_1) = \p_t A^{II}_{R,L}(k,t_1)$ \footnote{These matching conditions are related to the conservation of Wronskian for the propagating modes. } as
\bea
\big[ c^{s}_1 e^{-i\omega_{s}(k)t_1} - c^{s}_2 e^{i \omega_{s}(k) t_1}  \big] = \frac{\sqrt{k/2}}{\omega_{s}(k)} e^{-ikt_1},
\eea
which give
\bea\label{Cpm1}
c_1^{s} &=& \frac{1}{\sqrt{2k}}\frac{\omega_s(k)+k}{2\omega_s(k)} ~ e^{i(\omega_{s}(k)-k)t_1} ,\\ \label{Cpm2}
c_2^{s} &=& \frac{1}{\sqrt{2k}}\frac{\omega_s(k)-k}{2\omega_s(k)} ~ e^{-i(\omega_{s}(k)+k)t_1} .
\eea
ii) For $k < \xi:$ The minus helicity solution is given by Eq. \eqref{As} with $s=L$. However, the plus helicity solution takes the form
\bea
A_{R}(t,k) = \big[ c^{R}_1 e^{\lvert \omega_{R}(k)\rvert t} + c^{R}_2 e^{-\lvert \omega_{R}(k)\rvert t}  \big],
\eea
where  $c^{R}_{1,2} $  are given as \footnote{The conservation of the Wronskian demands $\lvert \omega_R(k)\rvert (c^{R*}_1c^{R}_2 -c^{R*}_2 c^{R}_1)=ik$.}
\bea\label{TCpm1}
c^{R}_1 &=& \frac{1}{\sqrt{2k}} \frac{i\lvert \omega_R(k)\rvert +k}{2i\lvert \omega_R(k)\rvert} ~ e^{-(\lvert \omega_{R}(k)\rvert +ik)t_1} ,\\ \label{TCpm2}
c^{R}_2 &=& \frac{1}{\sqrt{2k}} \frac{i\lvert \omega_R(k)\rvert -k}{2i\lvert \omega_R(k)\rvert} ~ e^{(\lvert \omega_{R}(k)\rvert -ik)t_1} .
\eea
{\bf{Region $III$ ($t\geq t_2$):}} The solution in region III can be written as
\bea
A^{III}_{s}(t,k) = \big[ d^{s}_1(k) e^{-ik t} + d^{s}_2(k) e^{ikt}  \big],
\eea
where $d^s_1(k)$ and $d^s_2(k)$ are specified by the matching conditions. i) For $k>\xi$, we find
\bea\label{ds1}
d^s_1(k) &=& \frac{1}{\sqrt{2k}}\bigg[ \frac{(\omega_s+k)^2}{4k\omega} e^{-i\omega_s(t_2-t_1)} - \frac{(\omega_s-k)^2}{4k\omega} e^{i\omega_s(t_2-t_1)} \bigg]e^{ik(t_2-t_1)} ,\\  \label{ds2}
d^s_2(k) &=& -\frac{i}{\sqrt{2k}}\frac{(k^2-\omega_s^2)}{2k\omega_s} \sin\omega_s(t_2-t_1) ~ e^{-ik(t_1+t_2)} .
\eea
iii) For $k< \xi$, the $d^-_1(k)$ and $d^-_1(k)$ coefficients are the same as Eq. \eqref{ds1} and \eqref{ds2}. The coefficients  of the plus helicity state are
\bea\label{ds1m}
d^+_1(k) &=& \bigg[ \frac{(i\lvert \omega_R \rvert+k)^2}{4i\lvert \omega_R \rvert k} e^{\lvert \omega_R \rvert(t_2-t_1)} - \frac{(i\lvert \omega_R \rvert-k)^2}{4i\lvert \omega_R \rvert k} e^{-\lvert \omega_R \rvert(t_2-t_1)}\bigg]e^{ik(t_2-t_1)} ,\\  \label{ds2m}
d^+_2(k) &=& \frac{(k^2+\lvert \omega_R \rvert^2)}{2i\lvert \omega_R \rvert k} \sinh\omega_R(t_2-t_1) ~ e^{-ik(t_1+t_2)} .
\eea